\documentclass[amsmath,reprint,prb,longbibliography,superscriptaddress,bibnotes]{revtex4-2}
\usepackage{amsthm,amssymb,amsfonts,graphicx,verbatim, xcolor,bm} 
\usepackage{hyperref} 
\usepackage[utf8]{inputenc} 
\usepackage{siunitx} 
\usepackage{braket}

\begin{document}

\title{Berry Phase Dynamics of Sliding Electron Crystals}
\author{Yongxin Zeng}
\email{yz4788@columbia.edu}
\affiliation{Department of Physics, Columbia University, New York, NY 10027}
\author{Andrew J. Millis}
\affiliation{Department of Physics, Columbia University, New York, NY 10027}
\affiliation{Center for Computational Quantum Physics, Flatiron Institute, New York, NY 10010}

\begin{abstract}
Systems such as Wigner crystals and incommensurate charge density waves that spontaneously break a continuous translation symmetry have unusual transport properties arising from their ability to slide coherently in space. Recent experimental and theoretical studies suggest that spontaneous translation symmetry breaking in some two-dimensional materials with nontrivial quantum geometry (e.g., rhombohedral pentalayer graphene) leads to a topologically nontrivial electron crystal state called the anomalous Hall crystal and characterized by a vanishing linear-response dc longitudinal conductivity and a non-vanishing Hall conductivity.  In this work we present a theoretical investigation of the sliding dynamics of this new type of electron crystal, taking into account the system's nontrivial quantum geometry. We find that when accelerated by an external electric field, the crystal acquires a transverse anomalous velocity that stems from not only the Berry curvature of the parent band but also the Galilean non-invariance of the crystal state (i.e., crystal states with different momenta are not related by simple momentum boosts). We further show that acceleration of the crystal modifies its internal current from the static crystal value that is determined by the Chern number of the crystal state. The net Hall conductance including contributions from center-of-mass motion and internal current is in general not quantized.
As an experimentally relevant example, we present numerical results in rhombohedral pentalayer graphene and discuss possible experimental implications.
\end{abstract}

\maketitle

\section{Introduction}

A gas of electrons in a uniform positive background interacting via the Coulomb potential undergoes spontaneous crystallization \cite{wigner1934crystal, tanatar1989ground} when the density is reduced below a critical value, breaking the continuous translational symmetry down to the discrete translational invariance of the crystal. In two dimensions, the ground state at low electron density is a triangular-lattice Wigner crystal (WC) whose properties are well understood \cite{bonsall1977static}. The recent experimental discoveries \cite{lu2024fractional, lu2024extended, waters2024interplay, xie2024even} of integer and fractional quantum anomalous Hall (QAH) effects in lightly doped rhombohedral pentalayer graphene was interpreted theoretically \cite{dong2023anomalous, dong2023theory, zhou2023fractional, guo2024fractional, kwan2023moire, zeng2024sublattice, tan2024parent, soejima2024anomalous, dong2024stability, zhou2024new} as arising from of a new type of electron crystal state called the anomalous Hall crystal (AHC). Unlike WCs, AHCs are honeycomb-lattice crystals in which electrons hop in a nontrivial way that leads to quantized Hall conductance \cite{haldane1988model}. While recent model studies \cite{zeng2024sublattice, tan2024parent, soejima2024anomalous, dong2024stability} have provided a qualitative understanding of the energetic competition between WCs and AHCs, the dynamical properties of AHCs remain unexplored. 

When a continuous translational symmetry is broken spontaneously, the energy of the symmetry-broken state is independent of its real-space location because of the translation-invariant background and the broken symmetry state can slide coherently in space. This phemonenon is well studied in the case of WCs \cite{glazman1992quantum, zhu1994nonlinear, zhu1994sliding, chitra1998dynamical, fogler2000dynamical} and incommensurate charge density waves (CDWs) \cite{peierls1996quantum, frohlich1954theory, lee1974conductivity, fukuyama1976pinning, fukuyama1978dynamics, lee2018electric, gruner1988cdw}. When the translational symmetry is explicitly broken by disorder and impurities in real materials, the CDW or WC is pinned at a local energy minimum until a large external electric field depins the crystallized electrons. The nonlinear current-voltage characteristics near the threshold voltage for the pinning-depinning transition are smoking gun signatures of spontaneous translational symmetry breaking.
Since AHCs also break continuous translational symmetry, we expect similar sliding transport phenomena. An important difference, however, is that the AHC is a topologically nontrivial state with chiral edge modes and quantized Hall conductance. It is natural to ask: Do AHCs go sideways when they slide?

If the AHC is treated classically, following the usual treatment of the sliding WC, translation invariance implies that in the absence of a magnetic field the crystal slides in the direction parallel to an applied force. On the other hand, because the parent band of the AHC usually contains a large amount of Berry flux, one may suspect \cite{patri2024extended} that a sliding AHC in an external electric field acquires a transverse anomalous velocity whose magnitude is determined by the average Berry curvature of the parent band weighted by the crystal state. In this paper we will show, via a detailed theoretical study of the sliding dynamics of electron crystals that takes into account Berry phase effects and applies to both WCs and AHCs, that this conclusion is not correct; the correct physical picture is much more subtle.

A central theme of our study is the violation of Galilean invariance. A translation invariant electron system is Galilean invariant when the physics in a moving reference frame described by the boosted Hamiltonian $H-\bm V\cdot\bm P$, where $\bm V$ is the velocity of the moving frame and $\bm P$ is the total momentum, is equivalent to that described by $H$ (up to constant momentum and energy shifts). This is the case for a single-component electron gas interacting via Coulomb potential. In a generic electron system, Galilean invariance is explicitly broken both by non-quadratic band dispersion and by the presence of nontrivial quantum geometry (i.e., nontrivial momentum dependence of cell-periodic Bloch wave functions). Broken Galilean invariance implies that sliding crystal states with different momenta are not related by simple momentum boosts. When a crystal accelerates (i.e., the total momentum varies with time), it adiabatically evolves into the ground state in a different reference frame. As a result, we show, an accelerating crystal acquires a transverse anomalous velocity that stems from not only the Berry curvature of the parent band but also crystal state deformation due to violation of Galilean invariance.

\begin{figure}
    \centering
    \includegraphics[width=\linewidth]{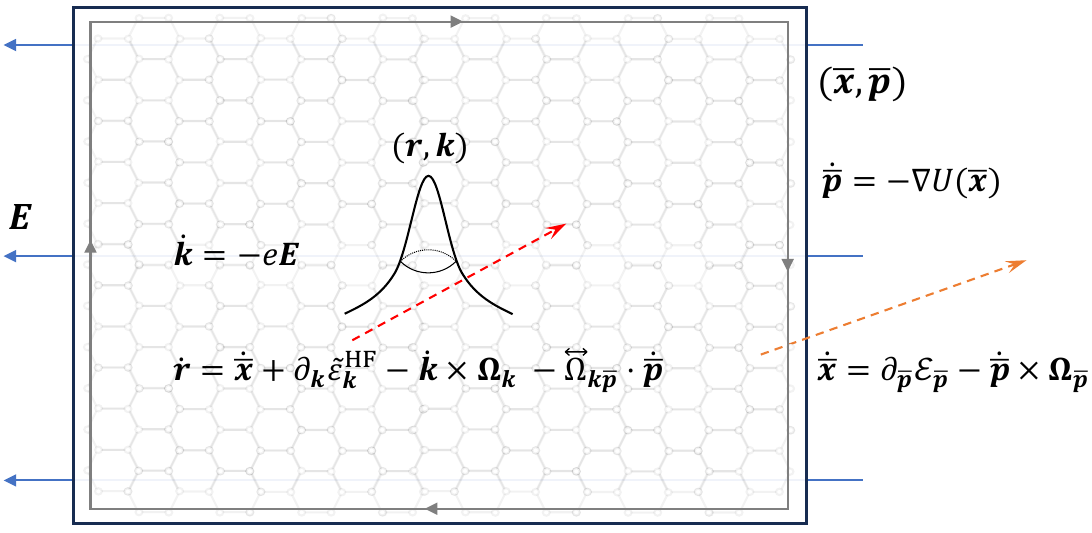}
    \caption{Schematic illustration of the contributions to the current in a sliding anomalous Hall crystal. The electron crystal is represented by the rectangular area with honeycomb-lattice pattern in light gray and hosts a chiral edge mode shown by the directed line circulating clockwise around the perimeter of the crystal arising from the nontrivial topology of the QAH state. The sliding crystal state is described by a many-body wave function $\ket{\Psi_{\bm{\bar x}, \bm{\bar p}}}$ (Eq.~\ref{eq:Psi_xp}) where $(\bm{\bar x}, \bm{\bar p})$ are the center-of-mass position and momentum (per electron) of the sliding crystal. $(\bm r, \bm k)$ are the position and momentum of a wave packet of Bloch electrons within the sliding crystal. Their equations of motion (see Sec.~\ref{sec:eom} for a derivation and precise definition of terms) are shown respectively on the right and inside the box representing the crystal. In an external electric field $\bm E$, the center-of-mass coordinates of the crystal accelerate and acquire a transverse anomalous velocity from the Berry phase derived term $\bm{\Omega_{\bar{p}}}$. Bloch electrons within the crystal also acquire an anomalous velocity from a combination of the Berry phase derived terms $\bm{\Omega_k}$ and $\tensor{\Omega}_{\bm{k\bar p}}$, the external electric field, and the acceleration of the crystal $\dot{\bm{\bar p}}$. }
    \label{fig:cartoon}
\end{figure}

The sliding motion of the electron crystal is not the only contribution to the total current; even in a pinned AHC, the motion of internal Bloch electrons leads to quantized Hall conductance due to the nontrivial topology of the crystal state. In a sliding electron crystal, the internal current is affected by the acceleration of the crystal and deviates from the quantized value given by the Chern number of the crystal state. A complete description of the dynamics of sliding crystals therefore requires two sets of equations, one for the center-of-mass sliding motion and the other for the motion of internal Bloch electrons.
The total current as the sum of the center-of-mass sliding current and the current of internal Bloch electrons is in general unquantized. The physical picture is schematically illustrated in Fig.~\ref{fig:cartoon}. While the general argument applies to both WCs and AHCs, we find from our numerical results that violation of Galilean invariance has a stronger effect on AHCs than on WCs.

The paper is organized as follows. In Sec.~\ref{sec:model} we introduce our model system and construct a group of sliding crystal states with different center-of-mass positions and momenta. In Sec.~\ref{sec:eom} we derive the equations of motion of the center-of-mass coordinates of the crystal and of the Bloch electrons within the crystal. As an experimentally relevant example, we present in Sec.~\ref{sec:rpg} numerical results in rhombohedral pentalayer graphene. Finally, we conclude in Sec.~\ref{sec:discussion} and discuss possible experimental implications of our results.

\section{Model} \label{sec:model}
Following Ref.~\cite{zeng2024sublattice}, we consider a two-dimensional system of spinless electrons with band dispersion $\varepsilon_{\bm p}$ and nontrivial quantum geometry encoded in the form factors $\Lambda_{\bm p', \bm p} = \braket{u_{\bm p'} | u_{\bm p}}$ where $\ket{u_{\bm p}}$ is the cell-periodic part of the Bloch wave function with momentum $\bm p$ in the microscopic lattice. A nontrivial quantum geometry corresponds to a nontrivial winding of $\ket{u_{\bm p}}$ in momentum space. The Hamiltonian of the system $H_0 = H_{\rm kin} + H_{\rm int}$ consists of kinetic energy
\begin{equation} \label{eq:H_kin}
H_{\rm kin} = \sum_{\bm p} \varepsilon_{\bm p} c_{\bm p}^{\dagger} c_{\bm p}
\end{equation}
and electron-electron interactions
\begin{equation} \label{eq:H_int}
H_{\rm int} = \frac{1}{2A} \sum_{\bm p \bm p' \bm q} V_{\bm q} \Lambda_{\bm p + \bm q, \bm p} \Lambda_{\bm p' - \bm q, \bm p'} c_{\bm p + \bm q}^{\dagger} c_{\bm p' - \bm q}^{\dagger} c_{\bm p'} c_{\bm p}.
\end{equation}
Here $c_{\bm p}^{\dagger}$ and $c_{\bm p}$ are the creation and annihilation operators of Bloch states with momentum $\bm p$, $V_{\bm q}$ is the Fourier transformed interaction potential, and $A$ is the area of the two-dimensional system. In our notation $\bm p$ represents momentum in the large Brillouin zone of the microscopic lattice, which for our purpose can be treated as infinite momentum space, and $\bm k$ (to appear below) represents crystal momentum within the mini Brillouin zone (mBZ) of the long-period crystal. Eqs.~\eqref{eq:H_kin}-\eqref{eq:H_int} describe the low-energy physics of interacting electrons in a multiband system projected onto the lowest-energy band; quantum geometry of the band enters the Hamiltonian as form factors that appear only in the interaction Hamiltonian $H_{\rm int}$ after band projection. The internal degrees of freedom including spin and valley are not written explicitly because it is assumed that these degrees of freedom are spontaneously polarized at low electron density, breaking time-reversal symmetry \footnote{An effective intra-spin-valley time-reversal operation can be defined and is a symmetry of the spin-valley-polarized system when $\varepsilon_{-\bm p} = \varepsilon_{\bm p}$ and $\Lambda_{-\bm p',-\bm p} = \Lambda_{\bm p', \bm p}^*$. See Ref.~\cite{zeng2024sublattice} for detailed discussion.}.

Eqs.~\eqref{eq:H_kin}-\eqref{eq:H_int} define a translationally invariant two-dimensional electron system. While the microscopic Hamiltonian has discrete translation symmetry due to the microscopic lattice potential, the effective low-energy model has an emergent continuous translation symmetry since the length scales of interest are much larger than that of the microscopic lattice. At low electron density, interactions dominate and the electron system crystallizes, spontaneously breaking the translational symmetry of the projected low-energy model. Two-dimensional electron crystals including WCs and AHCs form a long-period triangular Bravais lattice and correspondingly a short-period triangular reciprocal lattice. At the Hartree-Fock level, an electron crystal state is in general described by the Slater determinant
\begin{equation} \label{eq:Psi_0}
\ket{\Psi_0} = \prod_{\bm k \in {\rm mBZ}} \left[ \sum_{\bm g} v_{\bm k + \bm g} c_{\bm k + \bm g}^{\dagger} \right] \ket{0},
\end{equation}
where $\bm g$ represents the reciprocal lattice vectors of the long-period crystal. The coefficients $v$ are obtained by solving the self-consistent Hartree-Fock equations. As discussed below, the ground state is not unique; the real-space position of the self-consistent solution depends on the initial conditions of the numerical calculations.
Because of the nontrivial momentum dependence of the Bloch states created by $c_{\bm p}^{\dagger}$, the Berry connection $\bm{\mathcal{A}_k}$ contains the momentum derivatives of both the $v_{\bm p}$ coefficients and the $\ket{u_{\bm p}}$ states:
\begin{equation}
    \bm{\mathcal{A}_k} = \braket{v_{\bm k} | i\partial_{\bm k} | v_{\bm k}} + \sum_{\bm g} |v_{\bm k + \bm g}|^2 \braket{u_{\bm k + \bm g} | i\partial_{\bm k} | u_{\bm k + \bm g}}.
    \label{eq:A_k}
\end{equation}
From $\bm{\mathcal{A}_k}$ the Berry curvature $\bm{\Omega_k}$ and Chern number $C$ of the crystal state are determined as
\begin{align}
&\bm{\Omega_k} = \partial_{\bm k} \times \bm{\mathcal{A}_k}, \label{eq:Omega_k} \\
&C = \frac{1}{2\pi} \int_{\rm mBZ} d^2 \bm k\, \Omega_{\bm k}. \label{eq:C_k} 
\end{align}
Here the inner product notation of $v$-states represents an implicit summation over $\bm g$, e.g., $\braket{v_{\bm k} | i\partial_{\bm k} | v_{\bm k}} \equiv \sum_{\bm g} v_{\bm k + \bm g}^* i\partial_{\bm k} v_{\bm k + \bm g}$. Depending on the quantum geometry, band dispersion, and electron density, the ground state can be a $C=0$ WC or a $|C|=1$ AHC \cite{zeng2024sublattice}.

The translation invariance of the Hamiltonian Eqs.~\eqref{eq:H_kin}-\eqref{eq:H_int} means that $\ket{\Psi_0}$ described by Eq.~\eqref{eq:Psi_0} is not the unique ground state. Adding a momentum-dependent phase factor $v_{\bm p} \to v_{\bm p} e^{-i\bm p \cdot \bm x}$ shifts the crystal state by an in-plane vector $\bm x$ and results in an energetically equivalent ground state. The ground state degeneracy due to spontaneous breaking of continuous translation symmetry implies that the crystal state can evolve in a continuous family of many-body states with {\it collective} motion of all electrons.
The translation invariance also allows sliding crystal states with nonzero total momentum per particle $\bm{\bar{p}}$. A general sliding crystal state takes the form
\begin{equation} \label{eq:Psi_xp}
\ket{\Psi_{\bm{\bar x}, \bm{\bar p}}} = \prod_{\bm k \in {\rm mBZ}} \left[ \sum_{\bm g} v_{\bm k + \bm g}(\bm{\bar p}) e^{-i(\bm k + \bm g) \cdot \bm{\bar x}} c_{\bm k + \bm g}^{\dagger} \right] \ket{0}
\end{equation}
with $\bm{\bar p} \equiv \sum_{\bm p} |v_{\bm p}(\bm{\bar p})|^2 \bm p / N$ where $N$ is the total number of electrons. $\bm{\bar x}$ labels the real-space position of the crystal state; its precise definition and and  connection to electric polarization are discussed below. We call $(\bm{\bar x}, \bm{\bar p})$ the {\it center-of-mass} coordinates of the electron crystal despite the subtlety in defining the center of mass of a periodic system. Given $(\bm{\bar x}, \bm{\bar p})$, we assume that the ground state is unique and work under the adiabatic approximation such that the crystal state stays in the family of many-body states $\ket{\Psi_{\bm{\bar x}, \bm{\bar p}}}$ we construct.
We emphasize that $(\bm{\bar x}, \bm{\bar p})$ are the collective coordinates of a translation-symmetry-broken many-electron state rather than the coordinates of a single-electron wave packet, although their semiclassical equations of motion are derived in a similar way as for electron wave packets \cite{chang1996berry, sundaram1999wave} as detailed in the next section.

The coefficients $v_{\bm p}(\bm{\bar p})$ of a sliding crystal state are obtained by self-consistently solving the Hamiltonian in a moving frame $H_0 - \bm V \cdot \bm P$ where $\bm V$ is a constant velocity and $\bm P = \sum_{\bm p} \bm p c_{\bm p}^{\dagger} c_{\bm p}$ is the total momentum operator. In a Galilean-invariant system with quadratic dispersion $\varepsilon_{\bm p} = p^2/2m^*$ and trivial form factors $\Lambda_{\bm p', \bm p} = 1$, the Galilean-transformed Hamiltonian $H_0 - \bm V \cdot \bm P$ differs from $H_0$ only by a momentum shift $\bm{\bar p} = m^* \bm V$ (up to a trivial constant energy). The wave function of a sliding crystal in this case differs from that of a static crystal only by a momentum shift: $v_{\bm p}(\bm{\bar p}) = v_{\bm p - \bm{\bar p}}(0)$. In a general system with non-parabolic band dispersion and nontrivial form factors, e.g., rhombohedral multilayer graphene, Galilean invariance is lost and $v_{\bm p}(\bm{\bar p})$ with different $\bm{\bar p}$'s are not related by simple expressions.

While the momentum $\bm{\bar p}$ is well-defined once the coefficients $v_{\bm p}$ are known, the definition of the real-space coordinate of the crystal $\bm{\bar x}$ requires care because the {\it gauge transformation}
\begin{equation} \label{eq:gauge_trans}
v_{\bm p}(\bm{\bar p}) \to v_{\bm p}(\bm{\bar p}) e^{i\bm p \cdot \delta \bm x(\bm{\bar p})}, \quad \bm{\bar x} \to \bm{\bar x} + \delta \bm x(\bm{\bar p})
\end{equation}
shifts the origin of $\bm{\bar x}$ for each $\bm{\bar p}$ but leaves the crystal state \eqref{eq:Psi_xp} intact. Because sliding crystal states with different momenta $\bm{\bar p}$ are not related by simple boosts when Galilean invariance is broken, it is important to define in a consistent way the real-space position $\bm{\bar x}$ of sliding crystals with different momenta $\bm{\bar p}$. In other words, a definite phase choice is required for the coefficients $v_{\bm p}(\bm{\bar p})$. While the final results should be independent of the gauge choice (as we explicitly verify in Appendix~\ref{app:gauge_inv}), it is most natural to define the real-space position of the crystal state in terms of its electric polarization; otherwise the coupling to external scalar potential would be $\bm{\bar p}$-dependent.
The electric polarization of the state $\ket{\Psi_{\bm{\bar x}=0, \bm{\bar p}}}$ is given by the integral of Berry connection \cite{kingsmith1993theory, vanderbilt2018berry, xiao2010berry}:
\begin{align} \label{eq:polarization}
&\bm a_{\bm{\bar p}} = \frac{1}{N} \sum_{\bm k} \braket{\nu_{\bm k}(\bm{\bar p}) | i\partial_{\bm k} | \nu_{\bm k}(\bm{\bar p})} \notag \\
&= \frac{1}{N} \left[ \sum_{\bm k} \braket{v_{\bm k} | i\partial_{\bm k} | v_{\bm k}} + \sum_{\bm p} |v_{\bm p}|^2 \braket{u_{\bm p} | i\partial_{\bm p} | u_{\bm p}} \right],
\end{align}
where $\ket{\nu_{\bm k}(\bm{\bar p})} \equiv \sum_{\bm g} v_{\bm k + \bm g}(\bm{\bar p}) e^{i\bm g \cdot \bm r} \ket{u_{\bm k + \bm g}}$. A smooth gauge of Berry connection over the mBZ is required for an unambiguous definition of polarization (modulo lattice vectors of the crystal). For AHCs, additional care is needed because the integral of Berry connection for Chern insulators depends on the choice of origin of the mBZ \cite{coh2009electric}. The proper choice here, as shown in Appendix~\ref{app:polarization}, is a mBZ that is centered at $\bm{\bar p}$. Identification of the crystal position $\bm{\bar x}$ with polarization leads to the gauge-fixing condition $\bm{a_{\bar p}}=0$. Under this gauge, an adiabatic momentum boost of the crystal state does not shift the real-space position of the crystal. In any other gauge with nonzero $\bm{a_{\bar p}}$, the gauge-invariant crystal position is $\bm{\bar x} + \bm{a_{\bar p}}$. 
In the following sections we will fix the gauge such that $\bm{a_{\bar p}} = 0$ unless otherwise stated.


\section{Semiclassical dynamics of sliding crystals} \label{sec:eom}
In this section we derive the semiclassical equations of motion of sliding electron crystals. In Sec.~\ref{sec:com} we derive the equations of motion of the center-of-mass coordinates of the crystal, and in Sec.~\ref{sec:internal} we study the semiclassical dynamics of Bloch electrons within the crystal.

\subsection{Center-of-mass motion} \label{sec:com}
We start by considering a rigid electron crystal and leave a brief discussion of lattice deformations to the end of this section. The crystal state \eqref{eq:Psi_xp} is a semiclassical state characterized by position $\bm{\bar x}$ and momentum $\bm{\bar p}$. To study the dynamics of the sliding crystal, we consider the Hamiltonian $H = H_0 + \Delta H$ where $\Delta H$ describes a slowly varying external potential. By the time-dependent variational principles \cite{kramer1981geometry}, the semiclassical dynamics of the crystal state is described by the Lagrangian
\begin{align} \label{eq:L_x0p0}
L(&\bm{\bar x}, \bm{\bar p}, \dot{\bm{\bar x}}, \dot{\bm{\bar p}}) = \braket{\Psi_{\bm{\bar x}, \bm{\bar p}} | i\partial_t - H | \Psi_{\bm{\bar x}, \bm{\bar p}}} \notag \\
&= N \left[ \dot{\bm{\bar x}} \cdot \bm{\bar p} + \dot{\bm{\bar p}} \cdot \bm{\mathcal{A}}_{\bm{\bar p}} - \mathcal{E}_{\bm{\bar p}} - U(\bm{\bar x}) \right],
\end{align}
where the second line is obtained by using the explicit wave function \eqref{eq:Psi_xp} of the crystal state. The first two terms in the second line of Eq.~\eqref{eq:L_x0p0} arise from the time dependence of the crystal state through $\bm{\bar x}(t)$ and $\bm{\bar p}(t)$, where we have used the definition of $\bm{\bar p}$ below Eq.~\eqref{eq:Psi_xp} and the expression of $\bm{\mathcal{A}_{\bar p}}$ is provided in Eq.~\eqref{eq:Ap_short}, and the last two terms are the energy of the crystal state, specifically  $\mathcal{E}_{\bm{\bar p}} \equiv \braket{\Psi_{\bm{\bar x}, \bm{\bar p}} | H_0 | \Psi_{\bm{\bar x}, \bm{\bar p}}}/N$ is the average energy of electrons in the sliding crystal and $U(\bm{\bar x}) \equiv \braket{\Psi_{\bm{\bar x}, \bm{\bar p}} | \Delta H | \Psi_{\bm{\bar x}, \bm{\bar p}}}/N$ is the external potential acting on the crystal. We assume that $U$ is a smooth external potential that depends only on the polarization of the crystal $\bm{\bar x}$, e.g., in a uniform electric field $U(\bm{\bar x}) = e\bm E \cdot \bm{\bar x}$. The Berry connection for the center-of-mass motion $\bm{\mathcal{A}}_{\bm{\bar p}}$, related to but distinct from the Berry connection for a Bloch state $\bm{\mathcal{A}}_{\bm{k}}$ (Eq.~\ref{eq:A_k}), is given by
\begin{equation} \label{eq:Ap_short}
\bm{\mathcal{A}}_{\bm{\bar p}} = \frac 1N \sum_{\bm k} \braket{v_{\bm k}(\bm{\bar p}) | i\partial_{\bm{\bar p}} | v_{\bm k}(\bm{\bar p})}
\end{equation}
and describes the adiabatic evolution of the crystal state with the center-of-mass momentum of the crystal. With the gauge-fixing condition $\bm{a_{\bar p}}=0$ (see Eq.~\ref{eq:polarization}), the Berry connection is equivalently expressed as
\begin{align} \label{eq:Ap_long}
\bm{\mathcal{A}}_{\bm{\bar p}} = \frac{1}{N} \bigg[ &\sum_{\bm p} |v_{\bm p}(\bm{\bar p})|^2 \braket{u_{\bm p} | i\partial_{\bm p} | u_{\bm p}} \notag \\
&+ \sum_{\bm k} \braket{v_{\bm k}(\bm{\bar p}) | i(\partial_{\bm k} + \partial_{\bm{\bar p}}) | v_{\bm k}(\bm{\bar p})} \bigg].
\end{align}
The above expression provides intuition on the physical origin of the Berry connection $\bm{\mathcal{A}}_{\bm{\bar p}}$: the first term is the average Berry connection of Bloch states in the parent band weighted by occupation $|v_{\bm p}(\bm{\bar p})|^2$ and the second term arises from the Galilean non-invariance of the crystal state $v_{\bm p}(\bm{\bar p}) \ne v_{\bm p - \bm{\bar p}}(0)$.

The Euler-Lagrange equations
\begin{equation}
\frac{\partial L}{\partial \bm{\bar x}} = \frac{d}{dt} \frac{\partial L}{\partial \dot{\bm{\bar x}}}, \quad \frac{\partial L}{\partial \bm{\bar p}} = \frac{d}{dt} \frac{\partial L}{\partial \dot{\bm{\bar p}}}
\end{equation}
with the Lagrangian \eqref{eq:L_x0p0} lead to the equations of motion
\begin{align}
\dot{\bm{\bar x}} &= \partial_{\bm{\bar p}} \mathcal{E}_{\bm{\bar p}} - \dot{\bm{\bar p}} \times \bm \Omega_{\bm{\bar p}}, \label{eq:eom_x0} \\
\dot{\bm{\bar p}} &= -\nabla U(\bm{\bar x}), \label{eq:eom_p0}
\end{align}
where $\bm \Omega_{\bm{\bar p}} = \partial_{\bm{\bar p}} \times \bm{\mathcal{A}}_{\bm{\bar p}}$ is the Berry curvature for the center-of-mass motion of the crystal. In Appendix~\ref{app:symmetry} we show that $\bm \Omega_{\bm{\bar p}}$ has similar symmetry properties as the Berry curvature of Bloch bands: inversion symmetry implies $\bm \Omega_{-\bm{\bar p}} = \bm \Omega_{\bm{\bar p}}$ and time-reversal symmetry implies $\bm \Omega_{-\bm{\bar p}} = -\bm \Omega_{\bm{\bar p}}$.

Eqs.~\eqref{eq:eom_x0}-\eqref{eq:eom_p0} show that, similar to Bloch electrons moving in a periodic lattice potential, the crystal gains a transverse {\it anomalous velocity} when accelerated by an external force.
Despite the apparent similarity, Eqs.~\eqref{eq:eom_x0}-\eqref{eq:eom_p0} do not describe the motion of a single-electron wave packet in a periodic lattice potential, but rather the collective motion of a translation-symmetry-broken many-electron state in a translation-invariant background.
Because the Berry curvature $\bm \Omega_{\bm{\bar p}}$ arises from the center-of-mass momentum dependence of the crystal state rather than the Chern number of the crystal state, the anomalous velocity is nonzero for both AHCs and WCs as long as time-reversal symmetry is broken in the presence of nontrivial quantum geometry.

In a perfect translation-invariant system without disorder, the crystal state accelerates when a uniform electric field $\bm E$ is applied, leading to a transverse anomalous velocity that produces a Hall current $j_H = ne^2 \Omega_{\bm{\bar p}} E$ where $n=N/A$ is the electron density. The Hall conductivity $\sigma_{xy} = -2\pi n \Omega_{\bm{\bar p}} e^2/h$ is in general unquantized. In realistic systems, disorder and impurities provide a pinning potential and momentum-relaxation processes for the electron crystal. In a dc electric field, the electron crystal is pinned by impurities when the electric field is weak and moves at constant velocity when the electric field exceeds a threshold value. Assuming that disorder scattering acts as a smooth friction force that counteracts the electric field, the sliding electron crystal does not have an anomalous velocity. The realistic situation is complicated by extrinsic mechanisms of anomalous Hall effect \cite{nagaosa2010anomalous, sinitsyn2007semiclassical, xiao2010berry} including side jumps and skew scattering. As discussed in Sec.~\ref{sec:discussion}, we expect that a sliding electron crystal carries a nonzero Hall current associated with its center-of-mass motion and the net Hall conductance is in general unquantized.

To see the intrinsic effects of the anomalous velocity, we consider an ac electric field $\bm E(t) = E_0 e^{-i\omega t} \hat{\bm x}$ with frequency $\omega$ comparable to or higher than the disorder scattering rate. As a simple approximate model, we assume $\mathcal{E}_{\bm{\bar p}} \approx \bar p^2/2m$ and constant Berry curvature $\bm \Omega_{\bm{\bar p}} \approx \bm \Omega_0$ which are valid approximations at small $\bm{\bar p}$. We model the pinning potential as a simple harmonic potential $U_{\rm pin}(\bm{\bar x}) = m\omega_0^2 |\bm{\bar x}|^2 /2$ and model the effects of disorder scattering by a phenomenological damping term $-\bm{\bar p}/\tau$ with relaxation time $\tau$ \cite{gruner1988cdw, gruner1981nonlinear}. The equations of motion
\begin{align}
\dot{\bm{\bar x}} &= \frac{\bm{\bar p}}{m} - \dot{\bm{\bar p}} \times \bm \Omega_0, \\
\dot{\bm{\bar p}} &= -e\bm E - m\omega_0^2 \bm{\bar x} - \frac{\bm{\bar p}}{\tau}
\end{align}
have the steady-state solution with momentum
\begin{align}
\bar p_x &= \frac{(\omega_0^2 - \omega^2 - i\omega/\tau) i\omega eE}{(\omega_0^2 - \omega^2 - i\omega/\tau)^2 - (m\omega \omega_0^2 \Omega_0)^2}, \label{eq:px} \\
\bar p_y &= -\frac{m\omega^2 \omega_0^2 \Omega_0 eE}{(\omega_0^2 - \omega^2 - i\omega/\tau)^2 - (m\omega \omega_0^2 \Omega_0)^2}, \label{eq:py}
\end{align}
and corresponding real-space velocity
\begin{align}
\dot{\bar x} &= \frac{(\omega_0^2 - \omega^2 - i\omega/\tau - m^2 \omega^2 \omega_0^2 \Omega_0^2) i\omega eE/m}{(\omega_0^2 - \omega^2 - i\omega/\tau)^2 - (m\omega \omega_0^2 \Omega_0)^2}, \\
\quad \dot{\bar y} &= -\frac{(\omega + i/\tau) \omega^3 \Omega_0 eE}{(\omega_0^2 - \omega^2 - i\omega/\tau)^2 - (m\omega \omega_0^2 \Omega_0)^2}.
\end{align}
The oscillatory center-of-mass motion leads to longitudinal and transverse conductivity
\begin{align}
\sigma_{xx}^{\rm cm} &= -\frac{(\omega_0^2 - \omega^2 - i\omega/\tau - m^2 \omega^2 \omega_0^2 \Omega_0^2) i\omega ne^2/m}{(\omega_0^2 - \omega^2 - i\omega/\tau)^2 - (m\omega \omega_0^2 \Omega_0)^2}, \\
\sigma_{yx}^{\rm cm} &= \frac{(\omega + i/\tau) \omega^3 \Omega_0 ne^2}{(\omega_0^2 - \omega^2 - i\omega/\tau)^2 - (m\omega \omega_0^2 \Omega_0)^2}. \label{eq:sigma_cm_xy}
\end{align}
We see that the Berry curvature $\Omega_0$ leads to a transverse Hall current whose magnitude is strongly frequency-dependent. At low frequency $\omega\ll \omega_0, \tau^{-1}$, oscillation of the crystal is strongly damped and the Hall current increases with frequency as $\sim\omega^3$. At very high frequency $\omega\gg \omega_0, \tau^{-1}$, disorder and pinning effects are negligible and the Hall conductance saturates at $\sigma_{yx}^{\rm cm} \approx 2\pi n\Omega_0 e^2/h$.

So far we have focused on the case of a rigid electron crystal that stays perfectly periodic. In the more realistic case of deformable crystals, the center-of-mass coordinates of the crystal $(\bm{\bar x}, \bm{\bar p})$ should be promoted to the vector fields $(\bm X(\bm r), \bm{\Pi}(\bm r))$ that are defined as the local crystal displacement and momentum density. In an undeformed crystal, $\bm X = \bm{\bar x}$ and $\bm{\Pi} = n \bm{\bar p}$. The Lagrangian \eqref{eq:L_x0p0} is promoted to an integral
\begin{align}
L[\bm X(\bm r), \bm{\Pi}(\bm r)&] = \int d\bm r\, \Big[\dot{\bm{X}}(\bm r) \cdot \bm{\Pi}(\bm r) + \dot{\bm{\Pi}}(\bm r) \cdot \bm{\mathcal{A}}_{\bm{\Pi}(\bm r)/n} \notag \\
&-\frac{\bm{\Pi}^2(\bm r)}{2nm} - nU(\bm X(\bm r)) - \Delta \mathcal{E}(\partial\bm X) \Big],
\end{align}
where we have approximated the kinetic energy $\mathcal{E}_{\bm{\bar p}}$ by its leading quadratic term with effective mass $m$, and the weak spatial variation of local electron density is neglected since it is relevant only beyond quadratic order. The last term in the above expression represents the energy cost due to local lattice distortion:
\begin{align}
\Delta\mathcal{E} &= \frac{\kappa}{2}(\nabla\cdot\bm X(\bm r))^2 + \frac{\chi}{2}(\nabla\times\bm X(\bm r))^2 \notag \\
&+ \frac{n^2}{2} \int d\bm r'\, V(\bm r - \bm r') (\nabla\cdot\bm X(\bm r)) (\nabla\cdot\bm X(\bm r')),
\end{align}
where the first two terms describe short-range elastic energy with phenomenological constants $\kappa$ and $\chi$, and the last term describes long-range Coulomb interactions with $V(\bm r) = e^2/\epsilon r$. Assuming weak variation of Berry curvature $\bm{\Omega_{\bar p}} \approx \bm{\Omega}_0$, the equations of motion are
\begin{align}
\dot{\bm X}(\bm r) &= \frac{\bm{\Pi}(\bm r)}{nm} - \frac{\dot{\bm{\Pi}}(\bm r)}{n} \times \bm{\Omega}_0, \label{eq:X_dot} \\
\dot{\bm{\Pi}}(\bm r) &= -n \nabla U(\bm X(\bm r)) - \frac{\partial}{\partial \bm X} \Delta \mathcal{E}(\partial \bm X). \label{eq:Pi_dot}
\end{align}
The above equations of motion allow us to solve for the low-energy collective modes of the crystal. We find, as detailed in Appendix~\ref{app:phonon}, that the leading-order dispersion of the collective modes
\begin{equation} \label{eq:coll_mode}
\omega_+(q) \approx \left(\frac{2\pi e^2 n}{\epsilon m} q\right)^{1/2}, \quad \omega_-(q) \approx \left(\frac{\chi}{nm}\right)^{1/2} q
\end{equation}
are unaffected by the Berry curvature term. Here $\omega_+ \propto \sqrt{q}$ is the familiar longitudinal plasmon mode and $\omega_- \propto q$ is the transverse phonon mode. We find that the mass $m$ defined by the center-of-mass energy-momentum relation $\mathcal{E}_{\bm{\bar p}}$ is also the mass that determines the Drude weight and bulk plasmon dispersion of the sliding crystal. Despite the classical appearance of the plasmon dispersion $\omega_+(q)$, the effects of quantum geometry and broken Galilean invariance are implicitly included in the mass $m$. We note that broken Galilean invariance also underlies the quantum geometric contributions to the superfluid stiffness of superconductors \cite{zeng2025superfluid} and the Drude weight of Fermi liquids.

\subsection{Internal current} \label{sec:internal}
A QAH insulator has vanishing longitudinal conductivity $\sigma_{xx}=0$ and quantized Hall conductivity $\sigma_{xy} = -Ce^2/h$ where $C$ is the Chern number of the bulk band. Likewise, an AHC hosts an internal Hall current when an electric field is applied. Note that the internal current is separate from the center-of-mass motion and does not lead to a shift of charge density; the internal bulk current is compensated by unbalanced edge current and leads to vanishing net internal current \cite{vanderbilt2018berry}. When the crystal is pinned or sliding with constant velocity, the internal bulk current is a quantized Hall current given by the Chern number of the crystal state, similar to an ordinary QAH insulator. The total current is the sum of the sliding current and the internal current.

When the crystal accelerates, however, the internal current deviates from the quantized value. In a simple intuitive picture, electrons in the accelerating crystal frame feel an inertial force that counteracts the external electric field. In a Galilean non-invariant system with non-quadratic dispersion and nontrivial form factors, no simple expression exists for the inertial force and a careful wave packet construction \cite{chang1996berry, sundaram1999wave} is required to obtain the semiclassical equations of motion. Here we make use of the general results from Ref.~\cite{sundaram1999wave}, which in our context read
\begin{align}
\dot{\bm r} &= \partial_{\bm k} \varepsilon_{\bm k}^{\rm HF} - \tensor{\Omega}_{\bm k \bm k} \cdot \dot{\bm k} - \tensor{\Omega}_{\bm k \bm{\bar p}} \cdot \dot{\bm{\bar p}} - \tensor{\Omega}_{\bm k \bm{\bar x}} \cdot \dot{\bm{\bar x}}, \label{eq:rdot} \\
\dot{\bm k} &= -e\bm E.
\end{align}
Here $\varepsilon_{\bm k}^{\rm HF}$ is the Hartree-Fock energy of the state $\bm k$ in the lab frame, and $\tensor{\Omega}$'s are $2\times 2$ tensors with components
\begin{equation}
(\tensor{\Omega}_{\bm a \bm b})_{ij} = -i\,{\rm Im} \braket{\partial_{a_i} \nu | \partial_{b_j} \nu},
\end{equation}
where $\ket{\nu_{\bm k}(\bm{\bar x}, \bm{\bar p})} = \sum_{\bm g} v_{\bm k + \bm g}(\bm{\bar p}) e^{-i(\bm k + \bm g)\cdot\bm{\bar x}} e^{i\bm g \cdot \bm r} \ket{u_{\bm k + \bm g}}$ are Bloch states in the long-period crystal. In particular, $\tensor{\Omega}_{\bm k \bm k} \cdot \dot{\bm k} = \dot{\bm k} \times \bm{\Omega_k}$ where $\bm{\Omega_k} = \partial_{\bm k} \times \braket{\nu_{\bm k} | i\partial_{\bm k} | \nu_{\bm k}}$ is the Berry curvature of the Bloch states in the vector form. The last term in Eq.~\eqref{eq:rdot} is, after some algebra,
\begin{equation}
\tensor{\Omega}_{\bm k \bm{\bar x}} \cdot \dot{\bm{\bar x}} = \partial_{\bm k}(\dot{\bm{\bar x}} \cdot \bm p_{\bm k}) - \dot{\bm{\bar x}},
\end{equation}
where $\bm p_{\bm k} = \sum_{\bm g} |v_{\bm k + \bm g}|^2 (\bm k + \bm g)$ is the momentum of state $\ket{\nu_{\bm k}}$. This allows us to rewrite Eq.~\eqref{eq:rdot} in the form
\begin{equation} \label{eq:rdot_comov}
\dot{\bm r} = \dot{\bm{\bar x}} + \partial_{\bm k} \tilde{\varepsilon}_{\bm k}^{\rm HF} - \tensor{\Omega}_{\bm k \bm k} \cdot \dot{\bm k} - \tensor{\Omega}_{\bm k \bm{\bar p}} \cdot \dot{\bm{\bar p}},
\end{equation}
where $\tilde{\varepsilon}_{\bm k}^{\rm HF} = \varepsilon_{\bm k}^{\rm HF} - \dot{\bm{\bar x}} \cdot \bm p_{\bm k}$ is the Hartree-Fock energy of state $\ket{\nu_{\bm k}}$ in the co-moving frame of the crystal. In the form of Eq.~\eqref{eq:rdot_comov}, the sum of the first two terms on the right-hand side describe the group velocity of an electron wave packet in a moving crystal. The last term describes the modification of anomalous velocity due to acceleration (`inertial force') of the crystal. More explicitly,
\begin{equation} \label{eq:Okp_pdot}
-\tensor{\Omega}_{\bm k \bm{\bar p}} \cdot \dot{\bm{\bar p}} = (\dot{\bm{\bar p}} \cdot \partial_{\bm{\bar p}}) \braket{\nu | i\partial_{\bm k} | \nu} - \partial_{\bm k} \left[ \dot{\bm{\bar p}} \cdot \braket{\nu | i\partial_{\bm{\bar p}} | \nu} \right],
\end{equation}
where the first term arises from change of electric polarization due to change of center-of-mass momentum of the crystal and the second term describes an anomalous contribution to group velocity due to Berry phase accumulation of Bloch electrons in the accelerating crystal. A detailed derivation of Eq.~\eqref{eq:rdot_comov} based on wave packet construction is provided in Appendix~\ref{app:wave_packet}.


With the explicit expression of $\dot{\bm{\bar x}}$ in Eq.~\eqref{eq:eom_x0}, Eq.~\eqref{eq:rdot_comov} can be written as
\begin{equation}
\dot{\bm r} = \partial_{\bm{\bar p}} \mathcal{E}_{\bm{\bar p}} + \partial_{\bm k} \tilde{\varepsilon}_{\bm k}^{\rm HF} - \tensor{\Omega}_{\bm k \bm k} \cdot \dot{\bm k} - (\tensor{\Omega}_{\bm k \bm{\bar p}} + \tensor{\Omega}_{\bm{\bar p} \bm{\bar p}}) \cdot \dot{\bm{\bar p}}.
\end{equation}
The last term on the right describes the change of anomalous velocity due to acceleration of the crystal. Since $\tensor{\Omega}_{\bm k \bm{\bar p}} + \tensor{\Omega}_{\bm{\bar p} \bm{\bar p}} \ne 0$ in a system without Galilean invariance, the net Hall conductance is in general unquantized when the crystal accelerates.

Summing over all Bloch states in the crystal, in an external electric field $\bm E = E\hat{\bm x}$ the total internal current $\bm j^{\rm int} = -e \sum_{\bm k} (\dot{\bm r} - \dot{\bm{\bar x}})/A$ is
\begin{align}
j_x^{\rm int} &= \frac{e}{h} (C_{k_x \bar p_x} \dot{\bar p}_x + C_{k_x \bar p_y} \dot{\bar p}_y), \\
j_y^{\rm int} &= \frac{e^2}{h} CE + \frac{e}{h} (C_{k_y \bar p_x} \dot{\bar p}_x + C_{k_y \bar p_y} \dot{\bar p}_y),
\end{align}
where we defined the tensor
\begin{equation} \label{eq:C_tensor}
    \tensor{C}_{\bm{k\bar p}} = \frac{1}{2\pi} \int d^2 \bm k\, \tensor{\Omega}_{\bm{k\bar p}},
\end{equation}
which is formally similar to the Chern number $C$ (Eq.~\ref{eq:C_k}) but is in general unquantized when Galilean invariance is broken.

\section{Rhombohedral pentalayer graphene} \label{sec:rpg}
In this section we numerically calculate the energy-momentum relation and Berry curvatures of sliding crystals in the specific case of rhombohedral pentalayer graphene (PLG), the first system in which AHCs were proposed theoretically \cite{dong2023anomalous, dong2023theory, zhou2023fractional, guo2024fractional, kwan2023moire} as a possible explanation of the experimentally observed integer and fractional QAH effects \cite{lu2024fractional, lu2024extended}.

In the presence of a strong displacement field $u_D$ (defined as the potential difference between two adjacent layers; see Appendix~\ref{app:HF}), the first conduction band of PLG is energetically separated from the valence bands and higher conduction bands. Since the states we are interested in are spontaneously polarized in spin and valley, we model the system by the one-band Hamiltonian \eqref{eq:H_kin}-\eqref{eq:H_int} with $\varepsilon_{\bm p}$ and $\ket{u_{\bm p}}$ describing the dispersion and Bloch wave functions of the first conduction band of PLG in a single valley (see details in Appendix~\ref{app:HF}).

\begin{figure}
    \centering
    \includegraphics[width=0.8\linewidth]{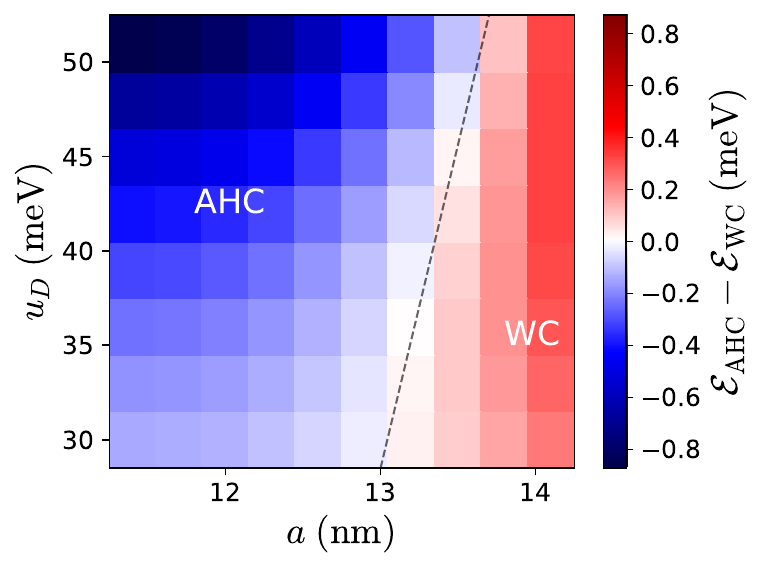}
    \caption{Energy difference (per electron) between static AHCs and WCs in rhombohedral pentalayer graphene. The AHC is the mean-field ground state in the blue region on the left of the dashed line and the WC is the ground state in the red region on the right.}
    \label{fig:dE}
\end{figure}

\begin{figure*}
    \centering
    \includegraphics[width=0.95\linewidth]{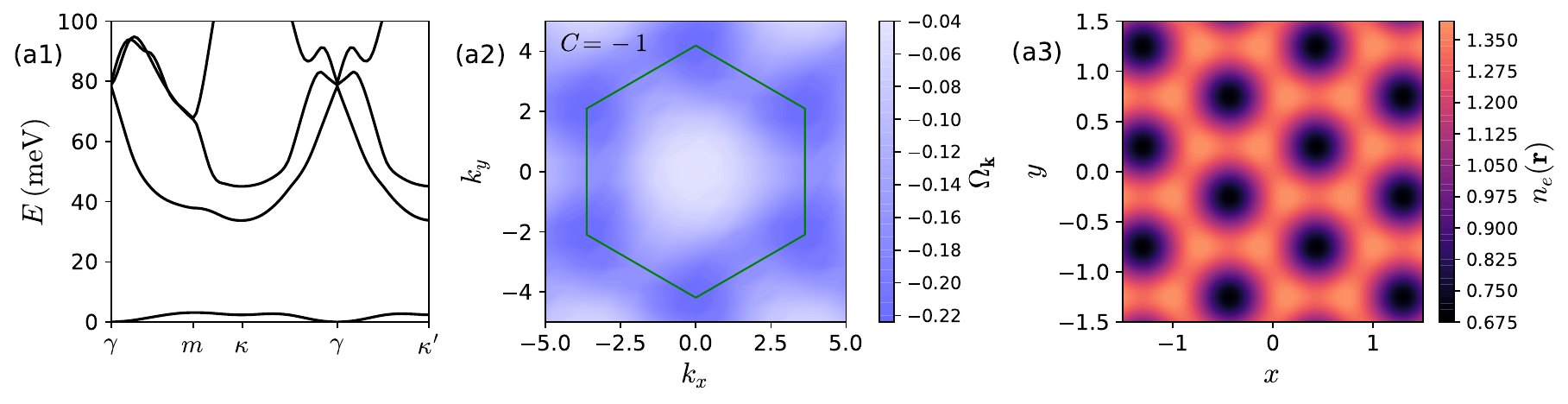}
    \includegraphics[width=0.95\linewidth]{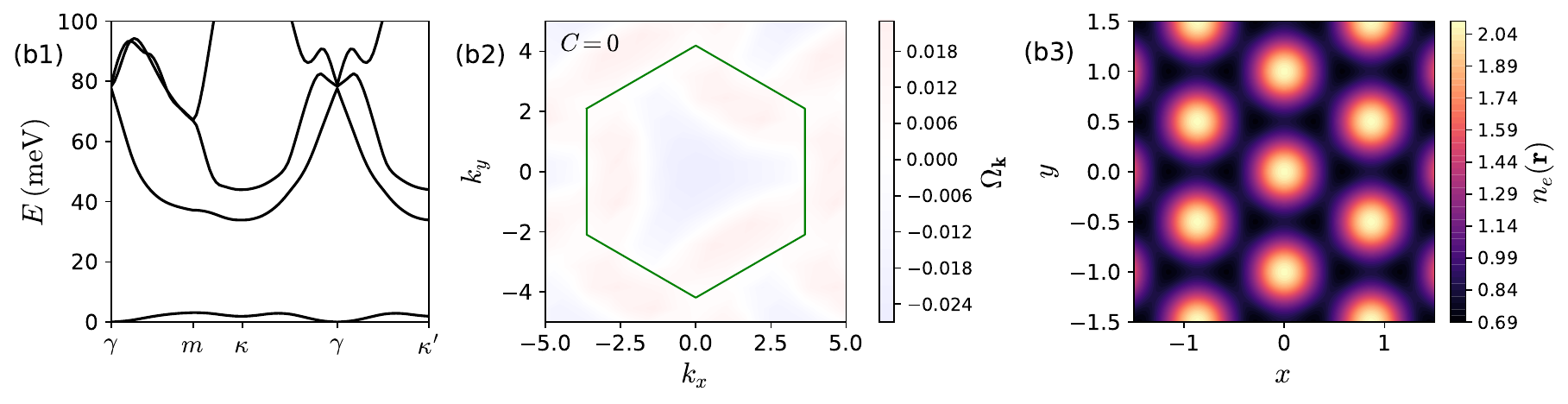}
    \includegraphics[width=0.95\linewidth]{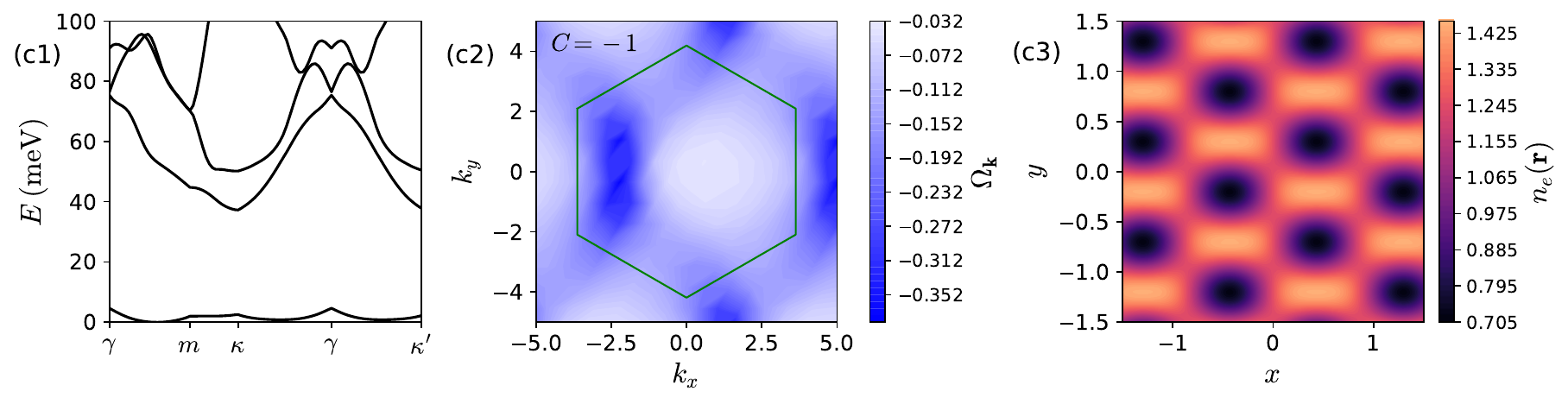}
    \includegraphics[width=0.95\linewidth]{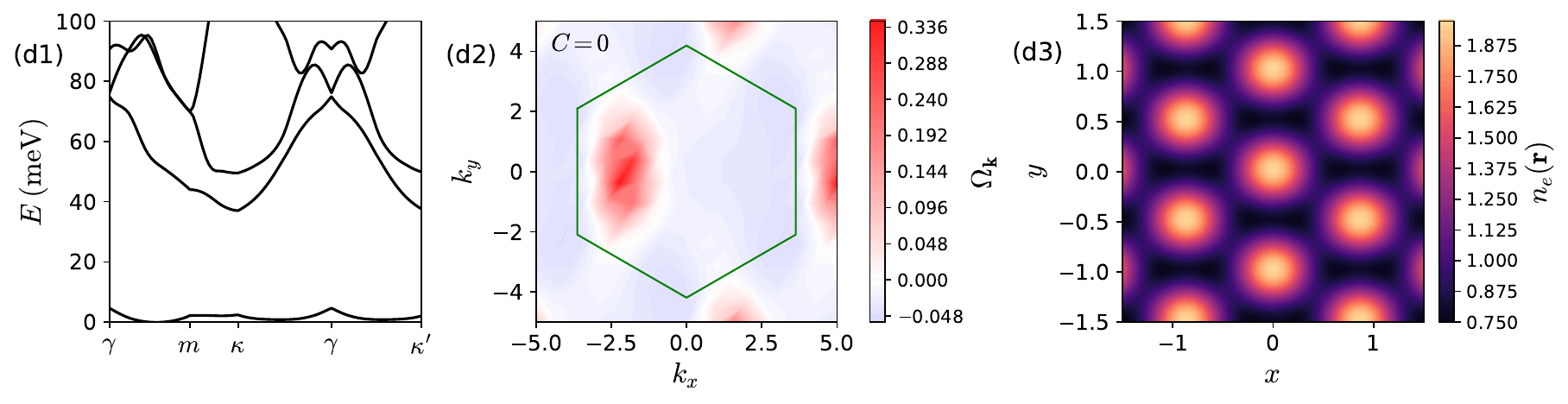}
    \caption{Electron crystal states in rhombohedral pentalayer graphene from self-consistent Hartree-Fock calculations. The displacement field $u_D = \SI{40}{meV}$ and the lattice constant $a = \SI{13}{nm}$. (a1-a3): Hartree-Fock band structure (with energy minimum set to zero), Berry curvature, and electron density of a static AHC; (b1-b3): same plots for a static WC; (c,d): same as (a,b) but the crystals slide in the $x$-direction with velocity $V_x = \SI{2e4}{m/s}$. The center-of-mass momenta (per electron) of the sliding crystals are (c) $\bar p_x = 1.300$; (d) $\bar p_x = 1.264$. Here all quantities are expressed with the length scale $a=1$.}
    \label{fig:crystals}
\end{figure*}

Given electron density $n$, we fold the band into the mBZ of a triangular lattice with lattice constant $a = (2/\sqrt{3}n)^{1/2}$ such that there is one electron per unit cell and perform self-consistent Hartree-Fock calculations. We find that for a wide range of parameters ($u_D$ and $a$), two types of crystal states that take the form of Eq.~\eqref{eq:Psi_0} exist as metastable self-consistent solutions of the Hartree-Fock equations. In Fig.~\ref{fig:dE} we plot the energy difference between two types of crystals and find that the WC is the mean-field ground state at weaker displacement fields and lower electron densities and the AHC is the ground state at stronger displacement fields and higher electron densities. Fig.~\ref{fig:crystals}(a-b) shows the mean-field band structure, Berry curvature $\Omega_{\bm k}$, and real-space electron density $n_e(\bm r)$ of two self-consistent crystal states at $u_D = \SI{40}{meV}$ and $a = \SI{13}{nm}$. While their mean-field band structures are almost identical, the two crystal states clearly have different sublattice structure and topology \cite{zeng2024sublattice}: the topologically nontrivial AHC (a) forms a honeycomb lattice and the topologically trivial WC (b) forms a triangular lattice.

The sliding crystal states are obtained by performing Hartree-Fock calculations with modified dispersion $\varepsilon_{\bm p}' = \varepsilon_{\bm p} - \bm V \cdot \bm p$ where $\bm V$ is the sliding velocity. The band structure, Berry curvature $\Omega_{\bm k}$, and charge density distributions in the co-moving frame of two crystal states sliding along the $x$-direction with velocity $V_x = \SI{2e4}{m/s}$ are shown in Fig.~\ref{fig:crystals}(c-d). While the sliding crystals have nonzero center-of-mass momenta, due to lack of Galilean invariance they are not related to the static crystals by simple momentum boosts. Our results show that both the Berry curvature and charge density distributions of the sliding crystals are different from those of the static crystals. Comparison of the charge density plots shows that the charge density of the sliding AHC is more strongly deformed than that of the sliding WC, suggesting that breaking of Galilean invariance has a stronger effect on AHCs than on WCs.

\begin{figure}
    \centering
    \includegraphics[width=0.75\linewidth]{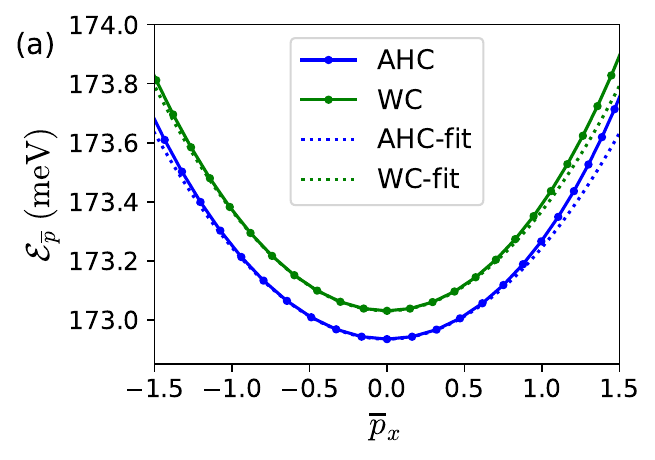}
    \includegraphics[width=\linewidth]{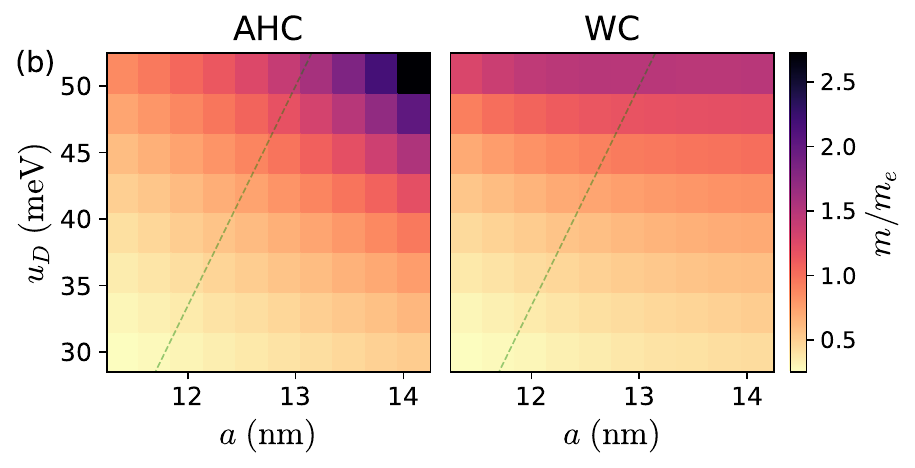}
    \caption{(a) Energy-momentum relation of sliding crystals at $u_D = \SI{40}{meV}$ and $a = \SI{13}{nm}$. The dotted curves are quadratic fitting of the numerical results with effective mass $m_{\rm AHC} = 0.73\,m_e$ and $m_{\rm WC} = 0.67\,m_e$. (b) Effective masses of the sliding crystals at varying $a$ and $u_D$. Two types of crystals have the same effective mass along the green dashed line; the AHC is heavier on the right and the WC is heavier on the left.}
    \label{fig:mass}
\end{figure}

Fig.~\ref{fig:mass}(a) shows the total energy -- center-of-mass momentum relation of two types of crystals sliding along the $x$-direction. While the electronic band dispersion of PLG is non-quadratic and anisotropic, $\mathcal{E}_{\bm{\bar p}}$ for both crystal states are approximately isotropic and quadratic at small $\bar p$. Anisotropy of the PLG band leads to slight asymmetry of the $\mathcal{E}_{\bar p_x}$ curve at large $\bar p_x$. The effective masses of the crystal states are obtained by quadratic fitting of the small-$\bar p$ part of the energy-momentum curve: $\mathcal{E}_{\bm{\bar p}} \approx \bar p^2/2m$. At $u_D = \SI{40}{meV}$ and $a = \SI{13}{nm}$, we find the effective mass of the AHC $m_{\rm AHC} = 0.73\,m_e$ is slightly larger than that of the WC $m_{\rm WC} = 0.67\,m_e$. Fig.~\ref{fig:mass}(b) shows the effective masses of two types of crystals in a range of parameters. The effective mass of the AHC increases rapidly with $a$ and $u_D$ and can be twice as large as that of the WC on the upper right corner of the plots. In contrast, the effective mass of the WC is less sensitive to $a$ and the WC is slightly heavier than the AHC at small $a$.

\begin{figure}
    \centering
    \includegraphics[width=0.9\linewidth]{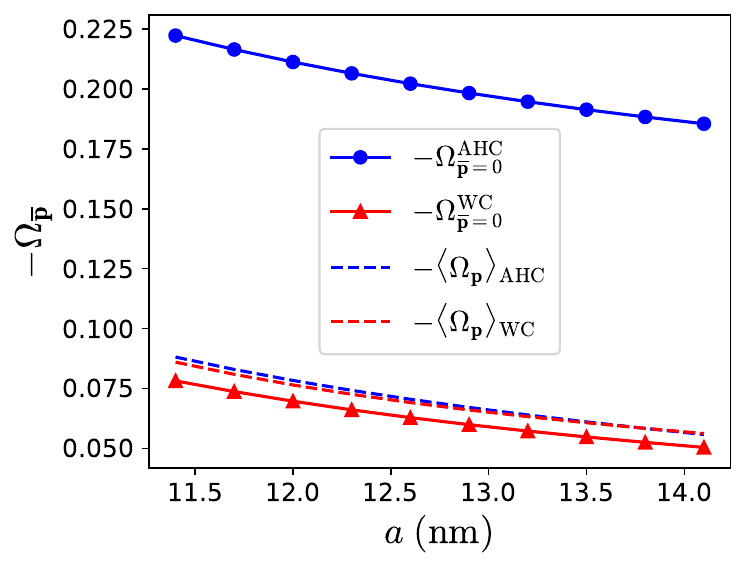}
    \caption{Berry curvature of the center-of-mass coordinates of the sliding crystals at $u_D = \SI{40}{meV}$. The solid lines with symbols show $\Omega_{\bm{\bar p}} = \partial_{\bm{\bar p}} \times \bm{\mathcal{A}_{\bar p}}$ (with $\bm{\mathcal{A}_{\bar p}}$ given by Eq.~\ref{eq:Ap_short}) computed at $\bm{\bar p}=0$. The dashed lines show the average Berry curvature of the parent band weighted by the static crystal states (Eq.~\ref{eq:Opp_av}).}
    \label{fig:Opp}
\end{figure}

Next we calculate the Berry curvature $\bm \Omega_{\bm{\bar p}} = \partial_{\bm{\bar p}} \times \bm{\mathcal{A}}_{\bm{\bar p}}$ of the sliding crystals with $\bm{\mathcal{A}}_{\bm{\bar p}}$ given by Eq.~\eqref{eq:Ap_short}. More specifically, we calculate the sliding crystal states around a small loop in $\bm{\bar p}$-space and calculate the Berry phase accumulated per unit area. The phases of the coefficients $v_{\bm p}(\bm{\bar p})$ must be adjusted by a gauge transformation \eqref{eq:gauge_trans} such that $\bm{a_{\bar p}} = 0$ (Eq.~\ref{eq:polarization}). We find that the $\bm{\bar p}$-dependence of $\Omega_{\bm{\bar p}}$ is negligibly weak for $\bar p \lesssim a^{-1}$, so in the following we focus on $\Omega_{\bm{\bar p}=0}$. The solid lines in Fig.~\ref{fig:Opp} show $\Omega_{\bm{\bar p}=0}$ for AHCs and WCs at fixed $u_D = \SI{40}{meV}$ and varying $a$ (the $u_D$-dependence of the results is relatively weak and is not important for our purpose). By comparison, the dashed lines show the average Berry curvature of the parent band weighted by the static crystal states:
\begin{equation} \label{eq:Opp_av}
\langle \Omega_{\bm p} \rangle = -\frac{1}{N} \sum_{\bm p} |v_{\bm p}|^2 \mathrm{Im} \braket{\partial_{p_x} u_{\bm p} | \partial_{p_y} u_{\bm p}}.
\end{equation}
We find that the band-averaged Berry curvature $\langle \Omega_{\bm p} \rangle$ for WCs and AHCs are almost indistinguishable, and that $\Omega_{\bm{\bar p}}$ for WCs is of similar magnitude as $\langle \Omega_{\bm p} \rangle$. In contrast, the magnitude of $\Omega_{\bm{\bar p}}$ for AHCs is about three times as large. Recall Eq.~\eqref{eq:Ap_long} where we showed that the difference between $\Omega_{\bm{\bar p}}$ and $\langle \Omega_{\bm p} \rangle$ arises from the Galilean non-invariance of the crystal states. The dramatic difference between $\Omega_{\bm{\bar p}}$ and $\langle \Omega_{\bm p} \rangle$ for AHCs suggests that violation of Galilean invariance has a stronger effect on AHCs than on WCs. This is consistent with the observation in Fig.~\ref{fig:crystals} that sliding AHCs have strongly deformed charge densities while sliding WCs are only slightly deformed.

\begin{figure}
    \centering
    \includegraphics[width=\linewidth]{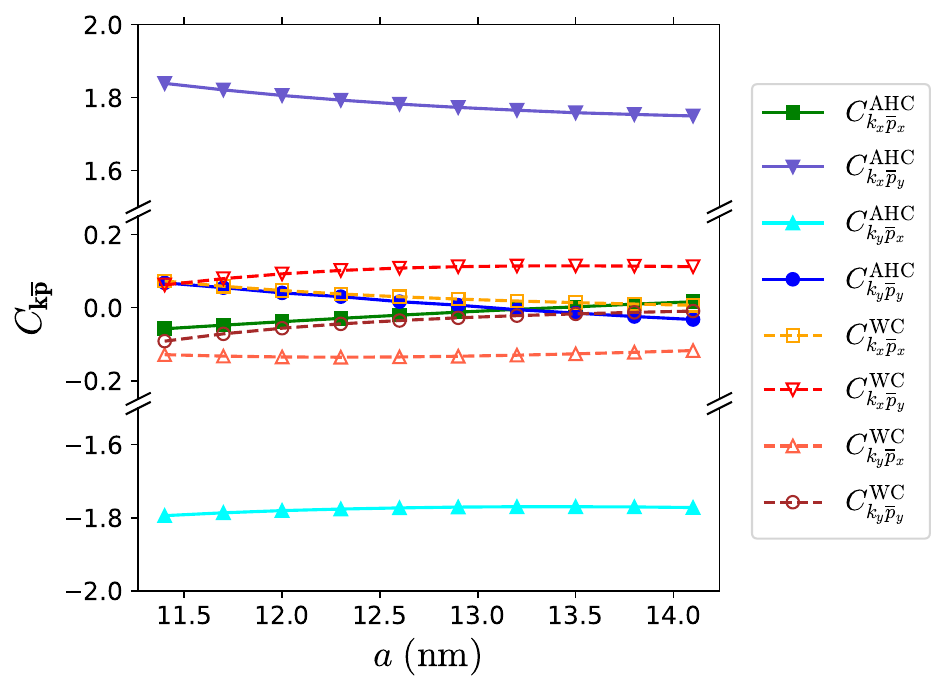}
    \caption{Components of the tensor $\tensor{C}_{\bm{k\bar p}}$ (Eq.~\ref{eq:C_tensor}) at $\bm{\bar p} = 0$ and $u_D = \SI{40}{meV}$. The results for AHCs are shown by solid lines with filled symbols and the results for WCs are shown by dashed lines with empty symbols.}
    \label{fig:Ckp}
\end{figure}

The components of the tensor $\tensor{C}_{\bm{k\bar p}}$ (Eq.~\ref{eq:C_tensor}) are similarly obtained by computing the Berry phase accumulated when the crystal state adiabatically evolves along a closed loop in $(k_i,\bar p_j)$ space. The results in Fig.~\ref{fig:Ckp} show that $\Omega_{k_x \bar p_y}$ and $\Omega_{k_y \bar p_x}$ for AHCs are much larger in magnitude than $\Omega_{k_x \bar p_x}$ and $\Omega_{k_y \bar p_y}$, which is expected because of the insulating bulk and nontrivial topology of AHCs. The precise values of $\Omega_{k_i \bar p_j}$ are, however, different from the $\Omega_{k_i \bar p_j} = -C\epsilon_{ij}$ expected in Galilean-invariant systems where acceleration of the crystal $\dot{\bar{\bm p}}$ produces an inertial force that acts as an effective electric field for the internal electrons. For WCs, all components of $\tensor{C}_{\bm{k\bar p}}$ are small but not precisely zero because of broken Galilean invariance.

\begin{figure}
    \centering
    \includegraphics[width=0.9\linewidth]{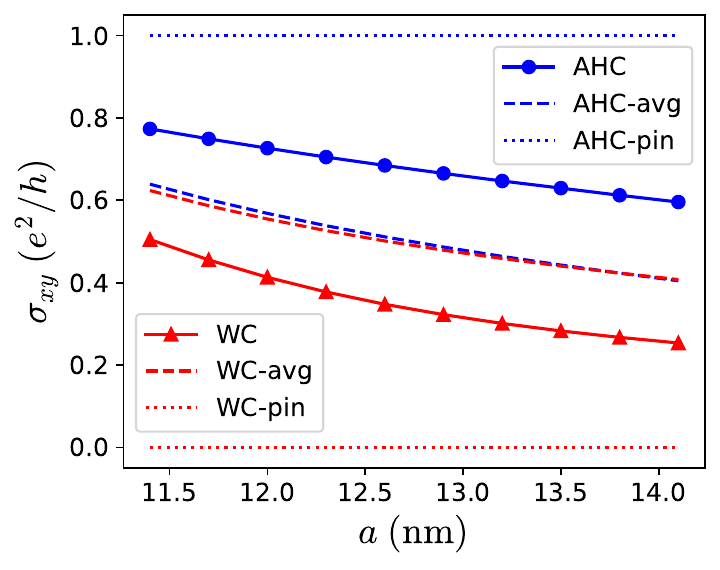}
    \caption{Net Hall conductance of sliding crystals at $u_D = \SI{40}{meV}$. The solid lines with symbols represent results for sliding crystals in the absence of pinning potential (Eq.~\eqref{eq:sxy}), the dashed lines represent results obtained by averaging the Berry curvature of the parent band (Eq.~\ref{eq:sxy_av}), and the dotted lines show results for pinned crystals in which case the Hall conductance is determined by the Chern number of the crystal state.}
    \label{fig:sxy}
\end{figure}

In the absence of pinning potential, $\dot{\bm k} = \dot{\bm{\bar p}} = -e\bm E$ and we get the net Hall conductance
\begin{equation} \label{eq:sxy}
\sigma_{xy} = -(C + C_{k_x \bar p_y} + 2\pi n \Omega_{\bm{\bar p}}) \frac{e^2}{h}.
\end{equation}
The results in Fig.~\ref{fig:sxy} show that the net Hall conductance of sliding AHCs and WCs are both between 0 and 1 in units of $e^2/h$, but the Hall conductance of sliding AHCs is about twice as large in magnitude as that of sliding WCs. The naive expectation that the Hall conductance is given by the average Berry curvature of the parent band
\begin{equation} \label{eq:sxy_av}
\sigma_{xy}^{\rm av} = -2\pi n \braket{\Omega_{\bm{\bar p}}} \frac{e^2}{h},
\end{equation}
with $\braket{\Omega_{\bm{\bar p}}}$ defined by Eq.~\eqref{eq:Opp_av}, gives nearly identical results for AHCs and WCs as shown by the dashed lines in Fig.~\ref{fig:sxy} that both fall between the correct $\sigma_{xy}$ for AHCs and WCs. Our results therefore suggest that a careful analysis of the Galilean non-invariant sliding crystal states is crucial to obtain the correct value of Hall conductance and other transport properties of sliding electron crystals.

\section{Discussion} \label{sec:discussion}
In this paper we theoretically studied the sliding dynamics of Wigner and anomalous Hall electron crystals moving in a parent band with nontrivial quantum geometry, focusing in particular on the effects of Berry curvature and broken Galilean invariance. We find that when accelerated by an external electric field, the center-of-mass motion of the crystal acquires a transverse anomalous velocity. Importantly, the relevant Berry curvature is not a simple average of the single-particle Berry curvature of the parent band as proposed by Ref.~\cite{patri2024extended}. Rather, the violation of Galilean invariance arising from the quantum geometry leads to an extra contribution that is dominant for AHCs in PLG as shown by our numerical results in Fig.~\ref{fig:Opp}. Acceleration of the crystal also affects the internal current produced by the external electric field, but its effect is again different from that of an inertial force on internal electrons due to violation of Galilean invariance. Fig.~\ref{fig:sxy} shows that the net Hall conductance falls between the value given by the average parent-band Berry curvature (Eq.~\ref{eq:sxy_av}) and the quantized value given by the Chern number of the crystal state.

While most of our theoretical discussion focused on the idealized case of electrons in a translation-invariant (non-disordered) background, an important consideration for experiment is the pinning potential due to disorder and impurities. When the electron crystal is pinned, the total current is completely determined by the topology of the crystal state; the WC has zero current while the AHC has a quantized Hall current. In a strong dc electric field, the crystal is depinned and moves with constant velocity. Naive application of our theory would suggest that the crystal without acceleration does not have a center-of-mass anomalous velocity, but this is not true because disorder scattering does not act as a smooth external potential that cancels out the electric field. Instead, disorder scattering events contribute to the anomalous Hall effect by {\it extrinsic} mechanisms \cite{nagaosa2010anomalous, xiao2010berry, sinitsyn2007semiclassical}. For example, a scattering event with transverse momentum shift $\Delta \bm{\bar p}$ leads to a longitudinal {\it side jump} $\Delta \bm{\bar x} = \bm{\Omega_{\bar p}} \times \Delta \bm{\bar p}$. Because side jumps along or against the electric field result in different electrostatic energies of the final states, the probabilities of scattering events in two transverse directions are different, leading to a net Hall current.

While a detailed study of the effects of pinning potential is left for future work, it is important to notice that because of broken Galilean invariance, any spatially non-uniform external potential is not just a function of the center-of-mass position of the crystal $\bm{\bar x}$ but also a function of momentum $\bm{\bar p}$. This is because the real-space charge density distribution changes with $\bm{\bar p}$ as shown in Fig.~\ref{fig:crystals}. The coupling between $\bm{\bar x}$ and $\bm{\bar p}$ in the external potential leads to a $\bm{\bar p}$-dependent external force and an $\bm{\bar x}$-dependent anomalous velocity that are largest when the external potential varies strongly on the length scale of the electron crystal $a$.

The recent experimental observation of integer QAH effect in an extended range of electron densities in PLG \cite{lu2024extended} suggests possible realization of AHCs incommensurate with the weak moir\'e potential. Remarkably, the dc transport data are consistent with a pinning-depinning transition, and the Hall conductance in the depinned regime is indeed unquantized. Another experimental system relevant to our theory is Bernal bilayer graphene in a strong displacement field where signatures of insulating states consistent with WCs or AHCs have been observed \cite{seiler2022quantum, seiler2024interaction}. More recently, signatures of depinning and sliding of electron crystals in bilayer graphene have been observed by current noise measurement \cite{seiler2024sliding}. Transport signatures of sliding WCs have also been observed in Cd$_3$As$_2$ thin films \cite{munyan2024evidence}, a strongly spin-orbit-coupled two-dimensional electron system. Further connection with our theory could be made in ac transport measurements in the pinned regime with frequency higher than or comparable to the momentum relaxation rate $\tau^{-1}$. Because the center-of-mass anomalous velocity is proportional to the acceleration by the electric field, we expect a frequency-dependent Hall conductance as in Eq.~\eqref{eq:sigma_cm_xy}.

The effective masses of sliding crystals we calculated in Fig.~\ref{fig:mass} determine the bulk plasmon dispersion \eqref{eq:coll_mode} that can be measured experimentally. The dependence of the plasmon dispersion on the carrier density in the extended QAH regime can distinguish the AHC and hole WC scenarios proposed in Ref.~\cite{patri2024extended}. Although the Berry curvature $\bm{\Omega_{\bar p}}$ does not enter the bulk plasmon dispersion, it gives rise to chiral plasmon modes at the edge \cite{song2016chiral}. Such {\it chiral Berry plasmons} have been measured experimentally in twisted bilayer graphene \cite{huang2022observation}, and we expect similar phenomena in sliding crystals in the presence of nonzero Berry curvature.

The difference in effective masses of different electron crystal states implies different dispersions of low-energy collective excitations which play an important role in the entropic competition between WCs, AHCs, and possibly fractional QAH crystals \cite{song2024intertwined, tan2024wavefunction}. The experimental observation \cite{lu2024extended} of extended integer QAH effect at the lowest temperatures and fractional QAH effect at slightly higher temperatures suggests that the integer QAH state (possibly AHC) has lower energy but the fractional QAH state has higher entropy \cite{patri2024extended, shavit2024entropy, wei2024edge}. If the fractional QAH state is a weakly pinned crystal state as proposed by Refs.~\cite{dong2023anomalous, dong2023theory, zhou2023fractional}, a careful examination of the sliding states and the associated collective modes of the integer and fractional QAH crystal states provides the key to resolving the experimental mysteries.

\begin{acknowledgements}
We thank Daniele Guerci, Daniel Mu\~noz-Segovia, Nemin Wei, Allan MacDonald, and Long Ju for helpful discussions.
Y.Z. and A.J.M. acknowledge support from Programmable Quantum Materials, an Energy Frontiers Research Center funded by the U.S. Department of Energy (DOE), Office of Science, Basic Energy Sciences (BES), under award DE-SC0019443.
This work was performed in part at the Aspen Center for Physics, which is supported by National Science Foundation grant PHY-2210452.
The Flatiron Institute is a division of the Simons Foundation.
\end{acknowledgements}

\appendix

\section{Polarization of sliding crystals} \label{app:polarization}

\begin{figure}
    \centering
    \includegraphics[width=\linewidth]{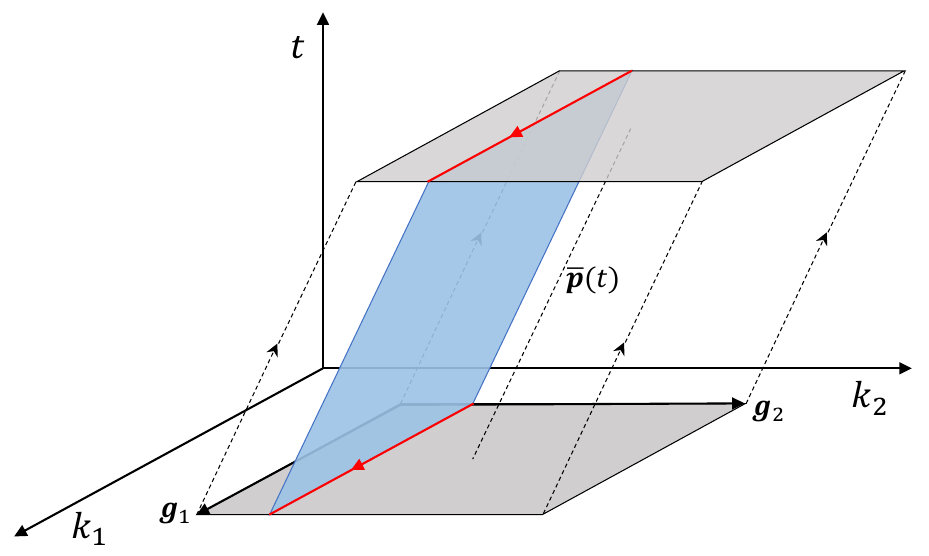}
    \caption{Adiabatic momentum boost of a crystal state. $k_1,k_2$ represent momenta along two reciprocal lattice vectors $\bm g_1, \bm g_2$, and the gray areas parallel to the $(k_1,k_2)$ plane represent the mBZ containing the Bloch states of the electron crystal at two times during the boost. The center of the mBZ follows (dotted line) the time evolution of the crystal momentum $\bm{\bar p}(t)$. The change of polarization along $k_1$ during the boost is the integral of Berry curvature $\Omega_{k_1 t}$ in the blue region, which up to integer multiples of lattice vectors is equal to the difference of Berry phases along $k_1$-loops at different $t$'s (red arrows, recall that the momenta are periodic in $\bm g$). The proper definition of the polarization is therefore the integral of the Berry connection (Eq.~\ref{eq:polarization}) in a smooth gauge over the mBZ centered at $\bm{\bar p}(t)$.}
    \label{fig:adiabatic_boost}
\end{figure}

To derive a well-defined polarization of the sliding crystal, we consider an adiabatic momentum boost of the crystal as shown in Fig.~\ref{fig:adiabatic_boost}. In the first-quantized form, the many-body wave function of the sliding crystal consists of a plane-wave factor $\exp(i\bm{\bar p} \cdot \sum_i \bm r_i)$ and a cell-periodic part
\begin{align}
\ket{U_{\bm{\bar p}}} &= \bigwedge_{\bm k} \sum_{\bm g} v_{\bm k + \bm g}(\bm{\bar p}) e^{i(\bm k + \bm g - \bm{\bar p}) \cdot \bm r} \ket{u_{\bm k + \bm g}} \notag \\
&= \bigwedge_{\bm k} \sum_{\bm g} v_{\bm k + \bm g + \bm{\bar p}}(\bm{\bar p}) e^{i(\bm k + \bm g) \cdot \bm r} \ket{u_{\bm k + \bm g + \bm{\bar p}}},
\end{align}
where $\bigwedge$ denotes the antisymmetric tensor product (Slater determinant) of single-particle states. The second expression makes it clear that the Bloch wave function of an electron in the sliding crystal evolves as
\begin{equation}
\ket{\nu_{\bm k + \bm{\bar p}(t)}(\bm{\bar p}(t))} = \sum_{\bm g} v_{\bm k + \bm g + \bm{\bar p}(t)}(\bm{\bar p}(t)) e^{i\bm g \cdot\bm r} \ket{u_{\bm k + \bm g + \bm{\bar p}(t)}}.
\end{equation}
When adiabatically boosted, up to first order in time derivative, the Bloch electron has velocity \cite{xiao2010berry, kingsmith1993theory, vanderbilt2018berry}
\begin{equation}
\bm{v_k}(t) = \partial_{\bm k} \varepsilon_{\bm k}^{\rm HF} - \Omega_{\bm k t},
\end{equation}
where the first term is the group velocity and the second term is a Berry curvature in $(\bm k, t)$-space defined as
\begin{equation}
\Omega_{\bm k t} \equiv -{\rm Im}\, \braket{\partial_{\bm k} \nu_{\bm k + \bm{\bar p}}(\bm{\bar p}) | \partial_t \nu_{\bm k + \bm{\bar p}}(\bm{\bar p})}.
\end{equation}
The change of polarization in the adiabatic process is given by the time integral
\begin{equation}
\Delta \bm a = \frac 1N \sum_{\bm k} \int dt\, \bm{v_k}(t) = -\frac 1N \sum_{\bm k} \int dt\, \Omega_{\bm k t}.
\end{equation}
Consider, for example, the component of $\Delta \bm a$ along one of the reciprocal lattice vectors $\bm g_1$. Given $k_2$, the above integral in $(k_1,t)$ space is shown schematically in Fig.~\ref{fig:adiabatic_boost} as the blue region which is a cylinder because of the periodicity along $\bm g_1$. The integral of curvature $\Omega_{k_1 t}$ in the blue region is equal to the difference of the integrals of Berry connection $\mathcal{A}_{k_1}(\bm{\bar p}) = \braket{\nu_{\bm k}(\bm{\bar p}) | i\partial_{k_1} | \nu_{\bm k}(\bm{\bar p})}$ along the two red arrows. More generally,
\begin{align}
\sum_{\bm k} \int_{t_1}^{t_2} dt\, \Omega_{\bm k t} = \sum_{\bm k} \big[&\bm{\mathcal{A}}_{\bm k + \bm{\bar p}(t_1)}(\bm{\bar p}(t_1)) \notag \\
&- \bm{\mathcal{A}}_{\bm k + \bm{\bar p}(t_2)}(\bm{\bar p}(t_2))\big].
\end{align}
A proper definition of the polarization of the sliding crystal is therefore
\begin{equation}
\bm{a_{\bar p}} = \frac 1N \sum_{\bm k} \braket{\nu_{\bm k + \bm{\bar p}}(\bm{\bar p}) | i\partial_{\bm k} | \nu_{\bm k + \bm{\bar p}}(\bm{\bar p})}
\end{equation}
with a fixed choice of mBZ (over which the integrand is smooth) for all $\bm{\bar p}$, or equivalently, Eq.~\eqref{eq:polarization} but with the mBZ centered at $\bm{\bar p}$.

\section{Symmetries of center-of-mass Berry curvature} \label{app:symmetry}
The center-of-mass Berry curvature is given by
\begin{equation}
\bm{\Omega}_{\bm{\bar p}} = \frac 1N \sum_{\bm k} \partial_{\bm{\bar p}} \times \braket{v_{\bm k}(\bm{\bar p}) | i\partial_{\bm{\bar p}} | v_{\bm k}(\bm{\bar p})}.
\end{equation}
Inversion symmetry guarantees $\varepsilon_{-\bm p} = \varepsilon_{\bm p}$ and $\Lambda_{-\bm p', -\bm p} = \Lambda_{\bm p', \bm p}$ in the Hamiltonian \eqref{eq:H_kin}-\eqref{eq:H_int}. This implies that the coefficients $v_{\bm p}(\bm{\bar p})$ for self-consistent mean-field solutions with opposite momenta are related by $v_{\bm p}(-\bm{\bar p}) = v_{-\bm p}(\bm{\bar p})$, and the Berry curvature satisfies
\begin{equation}
\bm{\Omega}_{-\bm{\bar p}} = \bm{\Omega_{\bar p}}.
\end{equation}
If time-reversal symmetry is preserved, $\varepsilon_{-\bm p} = \varepsilon_{\bm p}$ and $\Lambda_{-\bm p', -\bm p} = \Lambda_{\bm p', \bm p}^*$. This implies $v_{\bm p}(-\bm{\bar p}) = v_{-\bm p}^*(\bm{\bar p})$ and
\begin{equation}
\bm{\Omega}_{-\bm{\bar p}} = -\bm{\Omega_{\bar p}}.
\end{equation}
If inversion symmetry and time-reversal symmetry are both preserved, the Berry curvature $\bm{\Omega_{\bar p}}$ is identically zero for all $\bm{\bar p}$.

\section{Collective modes} \label{app:phonon}
In the absence of external potential, Eqs.~\eqref{eq:X_dot}-\eqref{eq:Pi_dot} read, in Fourier form
\begin{align}
-i\omega \bm X &= \frac{\bm{\Pi}}{nm} + \frac{i\omega}{n} \bm{\Pi}\times \bm{\Omega}_0, \\
-i\omega \bm{\Pi} &= -(n^2 V_q + \kappa) \bm q (\bm q \cdot \bm X) + \chi \bm q \times (\bm q \times \bm X),
\end{align}
where $V_q = 2\pi e^2/\epsilon q$ is the Fourier transformed Coulomb potential. Assuming $\bm q = q\hat{\bm x}$, the above equations can be written explicitly in components:
\begin{align}
-i\omega X &= \frac{\Pi_x}{nm} + \frac{i\omega}{n} \Omega_0 \Pi_y, \\
-i\omega Y &= \frac{\Pi_y}{nm} - \frac{i\omega}{n} \Omega_0 \Pi_x, \\
-i\omega \Pi_x &= -(2\pi e^2 n^2 q + \kappa q^2) X, \\
-i\omega \Pi_y &= -\chi q^2 Y.
\end{align}
Eliminating $\Pi_x, \Pi_y$ by simple substitution, we arrive at the secular equation
\begin{equation}
\begin{vmatrix}
\omega^2 - 2Dq - \frac{\kappa}{nm}q^2 & \frac{i\omega}{n}\Omega_0 \chi q^2 \\
i\omega \Omega_0 (2mDq + \frac{\kappa}{n}q^2) & \omega^2 - \frac{\chi}{nm}q^2
\end{vmatrix} = 0
\end{equation}
which determines the dispersion of collective modes. Here we defined the Drude weight $D = \pi e^2 n/\epsilon m$. The two solutions $\omega_{\pm}$ satisfy the relations
\begin{align}
\omega_+^2 + \omega_-^2 &= 2Dq + \frac{\kappa+\chi}{nm}q^2 - \frac{\Omega_0^2 \chi}{n^2} (2nmD + \kappa q) q^3, \\
\omega_+^2 \cdot \omega_-^2 &= \frac{2D\chi}{nm}q^3 + \frac{\kappa\chi}{n^2 m^2} q^4,
\end{align}
from which the leading-order expressions of $\omega_{\pm}(q)$ are readily obtained:
\begin{equation}
\omega_+(q) \approx (2Dq)^{1/2}, \quad \omega_-(q) \approx \left(\frac{\chi}{nm}\right)^{1/2} q.
\end{equation}
We find that because the Berry curvature $\Omega_0$ always appears with time derivatives of momenta, it does not enter the leading-order expressions of $\omega_{\pm}(q)$ but only shows up in higher-order expansions. However, this does not mean that quantum geometry has no effect on the Drude weight and bulk plasmon dispersion. Rather, quantum geometry affects the energy-momentum relation $\mathcal{E}_{\bm{\bar p}}$ and its effect on bulk plasmon dispersion is hidden in the mass $m$.

\section{Gauge invariance of the total current} \label{app:gauge_inv}
The physically measurable current in the sliding crystal $\bm j = -e\sum_{\bm k} \dot{\bm r}/A$ is the sum of center-of-mass current $\bm j^{\rm cm} = -ne\dot{\bm{\bar x}}$ and internal current $\bm j^{\rm int} = -e \sum_{\bm k} (\dot{\bm r} - \dot{\bm{\bar x}})/A$. The expressions of $\dot{\bm{\bar x}}$ and $\dot{\bm r}$ are given by Eqs.~\eqref{eq:eom_x0} and \eqref{eq:rdot_comov}. A gauge transformation $v_{\bm p}'(\bm{\bar p}) = v_{\bm p}(\bm{\bar p}) e^{i\bm p \cdot \delta \bm x(\bm{\bar p})}$ leads to a shift of center-of-mass origin $\bm{\bar x}' = \bm{\bar x} + \delta\bm x(\bm{\bar p})$. In a physically consistent theory, the sliding velocity must change by
\begin{equation} \label{eq:delta_xdot}
\dot{\bm{\bar x}'} = \dot{\bm{\bar x}} + (\dot{\bm{\bar p}} \cdot \partial_{\bm{\bar p}}) \delta \bm{\bar x},
\end{equation}
and the velocity of internal Bloch electrons in the crystal frame must change by an opposite amount, resulting in the same total velocity. Now we explicitly show these with the equations of motion derived in Sec.~\ref{sec:eom} and thus verify the consistency of our theory.

The $\bm{\bar p}$-dependent coordinate shift $\delta\bm x(\bm{\bar p})$ enters the Lagrangian \eqref{eq:L_x0p0} in two ways. First, it enters the external potential $U(\bm{\bar x}) = U(\bm{\bar x}' - \delta\bm x)$ and makes it $\bm{\bar p}$-dependent. This gives rise to an additional term in group velocity 
\begin{equation} \label{eq:dp_U}
\partial_{\bm{\bar p}} U = -\partial_{\bm{\bar p}} \delta\bm x \cdot \nabla U = \partial_{\bm{\bar p}} (\dot{\bm{\bar p}} \cdot \delta\bm x).
\end{equation}
In addition, $\delta\bm x(\bm{\bar p})$ enters the Berry connection \eqref{eq:Ap_short} as an extra term
\begin{equation}
\delta\bm{\mathcal{A}_{\bar p}} = \delta\bm x - \partial_{\bm{\bar p}} (\bm{\bar p} \cdot \delta\bm x)
\end{equation}
and modifies the anomalous velocity by
\begin{equation} \label{eq:delta_Omega_pdot}
(\partial_{\bm{\bar p}} \times \delta\bm{\mathcal{A}_{\bar p}}) \times \dot{\bm{\bar p}} = (\dot{\bm{\bar p}} \cdot \partial_{\bm{\bar p}}) \delta\bm x - \partial_{\bm{\bar p}} (\dot{\bm{\bar p}} \cdot \delta\bm x).
\end{equation}
The sum of Eqs.~\eqref{eq:dp_U} and \eqref{eq:delta_Omega_pdot} leads to the modified sliding velocity \eqref{eq:delta_xdot} as expected.

To see how the gauge transformation changes the velocity of internal Bloch electrons \eqref{eq:rdot_comov}, we first notice that the Hartree-Fock energy in the comoving frame $\tilde{\varepsilon}_{\bm k}^{\rm HF} = \varepsilon_{\bm k}^{\rm HF} - \dot{\bm{\bar x}} \cdot \bm p_{\bm k}$ changes by $\delta \tilde{\varepsilon}_{\bm k}^{\rm HF} = -\delta \dot{\bm{\bar x}} \cdot \bm p_{\bm k}$, resulting in a change of group velocity in the comoving frame
\begin{equation} \label{eq:dvg}
\partial_{\bm k} \delta \tilde{\varepsilon}_{\bm k}^{\rm HF} = -\partial_{\bm k} (\delta \dot{\bm{\bar x}} \cdot \bm p_{\bm k}).
\end{equation}
To see how $\delta\bm{\bar x}$ enters the last term in Eq.~\eqref{eq:rdot_comov}, we make use of Eq.~\eqref{eq:Okp_pdot} and find that
\begin{equation}
-\delta\tensor{\Omega}_{\bm k \bm{\bar p}} \cdot \dot{\bm{\bar p}} = -(\dot{\bm{\bar p}} \cdot \partial_{\bm{\bar p}}) \delta\bm{\bar x} + \partial_{\bm k} \left[ \bm{p_k} \cdot (\dot{\bm{\bar p}} \cdot \partial_{\bm{\bar p}}) \delta\bm{\bar x} \right].
\end{equation}
The last term cancels with Eq.~\eqref{eq:dvg}, leading to the net change of electron velocity in the crystal frame
\begin{equation}
\delta(\dot{\bm r} - \dot{\bm{\bar x}}) = -(\dot{\bm{\bar p}} \cdot \partial_{\bm{\bar p}}) \delta\bm{\bar x}.
\end{equation}
Together with Eq.~\eqref{eq:delta_xdot}, we have established the gauge invariance of the electron velocity in the lab frame and therefore the physically measurable total current.

\section{Wave packet dynamics of Bloch electrons in a sliding crystal} \label{app:wave_packet}
In this section we derive the semiclassical equations of motion of Bloch electrons using an explicit wave-packet construction that follows Refs.~\cite{sundaram1999wave, chang1996berry}. Consider a wave packet made from a linear superposition of Bloch states in a narrow range of $\bm k$:
\begin{align}
&\ket{W(t)} = \sum_{\bm k} w_{\bm k}(t) \ket{\nu_{\bm k}(\bm{\bar x}(t),\bm{\bar p}(t))} \notag \\
&=\sum_{\bm k} w_{\bm k}(t) \sum_{\bm g} v_{\bm k + \bm g}(\bm{\bar p}(t)) e^{-i(\bm k + \bm g) \cdot \bm{\bar x}(t)} c_{\bm k + \bm g}^{\dagger} \ket{0}.
\end{align}
The wave packet has a distribution in $\bm k$-space centered at
\begin{equation}
\bm k_c(t) \equiv \sum_{\bm k} \bm k |w_{\bm k}(t)|^2.
\end{equation}
In real space, the wave packet is centered at
\begin{equation}
\bm r_c(t) \equiv \braket{W(t) | \bm r | W(t)} = \sum_{\bm k} w_{\bm k}^* i\partial_{\bm k} w_{\bm k} + \bm{\mathcal{A}}_{\bm k_c} + \bm{\bar x},
\end{equation}
where we assumed a narrow distribution in $\bm k$ such that $|w_{\bm k}|^2 \approx \delta(\bm k - \bm k_c)$.
The semiclassical dynamics of the wave packet is described by the Lagrangian
\begin{align}
&L(\bm r_c, \bm k_c, \dot{\bm r}_c, \dot{\bm k}_c) = \braket{W(t) | i\partial_t - H | W(t)} \notag \\
= &(\dot{\bm r}_c - \dot{\bm{\bar x}}) \cdot \bm k_c + \dot{\bm k}_c \cdot \bm{\mathcal{A}}_{\bm k_c} + \dot{\bm{\bar p}} \cdot \bm{\mathcal{A}_{\bar p}}(\bm k_c) \notag \\
&+ \dot{\bm{\bar x}} \cdot \bm p_{\bm k_c} - \varepsilon_{\bm k_c}^{\rm HF} - e\bm E \cdot \bm r_c,
\end{align}
where total time derivative terms are dropped and we have defined
\begin{equation}
\bm{\mathcal{A}}_{\bm{\bar p}}(\bm k) \equiv \braket{v_{\bm k}(\bm{\bar p}) | i\partial_{\bm{\bar p}} | v_{\bm k}(\bm{\bar p})}.
\end{equation}
It follows that the coordinates of the wave packet fulfill the equations of motion
\begin{align}
\dot{\bm k}_c &= -e\bm E, \\
\dot{\bm r}_c &= \dot{\bm{\bar x}} + (\dot{\bm k}_c \cdot \partial_{\bm k_c} + \dot{\bm{\bar p}} \cdot \partial_{\bm{\bar p}}) \bm{\mathcal{A}}_{\bm k_c} \notag \\
&+ \partial_{\bm k_c} (\tilde{\varepsilon}_{\bm k_c}^{\rm HF} - \dot{\bm{\bar p}} \cdot \bm{\mathcal{A}_{\bar p}}(\bm k_c) - \dot{\bm k}_c \cdot \bm{\mathcal{A}}_{\bm k_c}), \label{eq:rc_dot}
\end{align}
where $\tilde{\varepsilon}_{\bm k}^{\rm HF} \equiv \varepsilon_{\bm k}^{\rm HF} - \dot{\bm{\bar x}} \cdot \bm{p_k}$ is the mean-field energy of Bloch states in the crystal frame. Using Eq.~\eqref{eq:Okp_pdot} and
\begin{equation}
\dot{\bm k}_c \times (\partial_{\bm k_c} \times \bm{\mathcal{A}}_{\bm k_c}) = \partial_{\bm k_c} (\dot{\bm k}_c \cdot \bm{\mathcal{A}}_{\bm k_c}) - (\dot{\bm k}_c \cdot \partial_{\bm k_c}) \bm{\mathcal{A}}_{\bm k_c},
\end{equation}
Eq.~\eqref{eq:rc_dot} becomes
\begin{equation}
\dot{\bm r}_c = \dot{\bm{\bar x}} + \partial_{\bm k_c} \tilde{\varepsilon}_{\bm k_c}^{\rm HF} - \dot{\bm k}_c \times \bm{\Omega}_{\bm k_c} - \tensor{\bm{\Omega}}_{\bm k_c \bm{\bar p}} \cdot \dot{\bm{\bar p}},
\end{equation}
in agreement with Eq.~\eqref{eq:rdot_comov}.

\section{Hartree-Fock calculations in PLG} \label{app:HF}
\subsection{Rhombohedral pentalayer graphene (PLG)}
The single-particle Hamiltonian of PLG is a $10\times 10$ matrix in layer-sublattice space. In layer space,
\begin{equation} \label{eq:H_PLG}
H_{\mathrm{PLG}} = \begin{pmatrix}
h_0+2u_D & h_1 & h_2 & 0 & 0 \\
h_1^{\dagger} & h_0+u_D & h_1 & h_2 & 0 \\
h_2^{\dagger} & h_1^{\dagger} & h_0 & h_1 & h_2 \\
0 & h_2^{\dagger} & h_1^{\dagger} & h_0-u_D & h_1 \\
0 & 0 & h_2^{\dagger} & h_1^{\dagger} & h_0-2u_D
\end{pmatrix},
\end{equation}
where each $h$ is a $2\times 2$ matrix in sublattice space. The diagonal blocks
\begin{equation}
h_0(\bm p) = \begin{pmatrix}
0 & v_0 p_- \\
v_0 p_+ & 0
\end{pmatrix},
\end{equation}
where $p_{\pm} = \chi_v p_x \pm ip_y$ with $\chi_v=\pm$ depending on the valley (we choose $\chi_v=+$ in our single-valley single-spin calculations), describe the low-energy physics of Dirac electrons in each graphene layer. The off-diagonal blocks
\begin{equation}
h_1(\bm p) = \begin{pmatrix}
v_4 p_- & v_3 p_+ \\
\gamma_1 & v_4 p_-
\end{pmatrix}, \quad
h_2 = \begin{pmatrix}
0 & \gamma_2/2 \\
0 & 0
\end{pmatrix}
\end{equation}
describe hopping between neighboring and next-neighboring layers respectively. The effect of an applied displacement field is modelled as a potential difference $u_D$ between neighboring layers. We use parameters from Ref.~\cite{zhou2021half} which are chosen to match the experimental data in rhombohedral trilayer graphene: $v_0 = \SI{1.0e6}{m/s}$, $\gamma_1 = \SI{380}{meV}$, $\gamma_2 = \SI{-15}{meV}$, $v_3 = \SI{-9.4e4}{m/s}$, $v_4 = \SI{-4.6e4}{m/s}$.

\subsection{Hartree-Fock calculations}
The band structure of PLG is obtained by solving for the eigenvalues and eigenstates of Eq.~\eqref{eq:H_PLG}. For our purpose, the only relevant band is the first conduction band with dispersion $\varepsilon_{\bm p}$ and Bloch eigenstates $\ket{u_{\bm p}}$. Projecting to the first conduction band, we get the Hamiltonian \eqref{eq:H_kin}-\eqref{eq:H_int} with form factors $\Lambda_{\bm p', \bm p} = \braket{u_{\bm p'} | u_{\bm p}}$. For the interaction potential, we use gate screened Coulomb potential $V_{\bm q} = (2\pi e^2/\epsilon q) \tanh(qd)$ with dielectric constant $\epsilon=5$ and distance to gate $d=\SI{20}{nm}$. To get the electron crystal states, we fold the band into a mBZ with reciprocal lattice vectors $\bm g_1 = (4\pi/\sqrt{3}a,0)$, $\bm g_2 = (2\pi/\sqrt{3}a,2\pi/a)$ where $a = (2/\sqrt{3}n)^{1/2}$ is the lattice constant determined by the electron density $n$. The mean-field Hamiltonian is $H_{\rm MF} = H_{\rm kin} + \Sigma_{\rm H} + \Sigma_{\rm F}$ with the Hartree term
\begin{equation}
\Sigma_{\rm H} = \frac 1A \sum_{\bm p \bm p' \bm g} V_{\bm g} \Lambda_{\bm p+\bm g,\bm p} \Lambda_{\bm p'-\bm g,\bm p'} \braket{c_{\bm p'-\bm g}^{\dagger} c_{\bm p'}} c_{\bm p+\bm g}^{\dagger} c_{\bm p}
\end{equation}
and the Fock term
\begin{equation}
\Sigma_{\rm F} = -\frac 1A \sum_{\bm p \bm p' \bm g} V_{\bm p' - \bm p - \bm g} \Lambda_{\bm p'-\bm g,\bm p} \Lambda_{\bm p+\bm g,\bm p'} \braket{c_{\bm p'-\bm g}^{\dagger} c_{\bm p'}} c_{\bm p+\bm g}^{\dagger} c_{\bm p}.
\end{equation}
The electron crystal states are obtained by solving the Hartree-Fock equations self-consistently. In our numerical calculations we use a $9\times 9$ $\bm k$-grid in the mBZ and three to four $\bm g$-shells that ensure convergence of our results.

\bibliography{ref}
\end{document}